\documentclass[12pt, letter]{article}

\usepackage{graphicx}
\usepackage{amsmath, amssymb,theorem,verbatim}
\usepackage{apacite}
\usepackage{textcomp} 
\usepackage{setspace}
\usepackage{subfigure}  
\usepackage{multirow} 
\usepackage{rotating}

\long\def\comment#1{}

\newcommand{\Ex}{\mathbb{E}}
\newcommand{\cov}{\textrm{cov}}
\newcommand{\var}{\textrm{var}}
\newcommand{\cum}{\textrm{cum}}
\newcommand{\Dcon}{\stackrel{D}{\rightarrow}}
\newcommand{\Pcon}{\stackrel{P}{\rightarrow}}

\newtheorem{theorem}{Theorem}[section]

\newtheorem{lemma}{Lemma}[section]

\newtheorem{assumption}{Assumption}[section]

\newtheorem{remark}{Remark}[section]

\newtheorem{corollary}{Corollary}[section]

\begin{document}

\title{A test for second order stationarity 
of a time series based on the
Discrete Fourier Transform (Technical Report)}
\author{Yogesh Dwivedi and Suhasini Subba Rao \\
Texas A\&M  University, \\
College Station, Texas 77845, USA}


\maketitle

\begin{abstract}
 We consider a zero mean discrete time series, and define its discrete Fourier
transform at the canonical frequencies. It is well known that the 
discrete Fourier transform is asymptotically uncorrelated at the 
canonical frequencies if and if only the time series is second order
stationary. Exploiting this important property, we construct a Portmanteau
type test statistic for testing stationarity of the time series. 
It is shown that under the null of stationarity, the test statistic is 
approximately a chi square distribution. 
To examine the power of the test statistic, the asymptotic distribution 
under the  locally stationary alternative is established. It is shown to be 
a type of noncentral chi-square, where the noncentrality parameter  
measures the deviation from stationarity. 
The test is illustrated with simulations, where is it shown to have 
good power.
Some real examples are also included to illustrate the test.

{\bf Keywords and phrases} Discrete Fourier Transform, 
linear time series, local stationarity, Portmanteau test, 
test for second order stationarity. 

\end{abstract}

\section{Introduction}

An important assumption that is often made when analysing time series is
that it is at least second order stationary. Most of the linear time
series literature is based on this assumption. If the assumption is not
properly tested and the analysis is performed, the resulting models are
considered to be misspecified and any forecasts obtained are not
appropriate. Therefore, it is important to check whether
the time series is second order stationary.

In recent years several statistical tests have been proposed. 
Many of the proposed tests
are based on comparing spectral densities over various segments 
(\citeA{p:pri-sub-69}, \citeA{p:pic-85}, \citeA{p:gir-lep-92}, \citeA{p:ada-98} and \citeA{p:pap-09})
comparing covariance structures over various
segments of the data (\citeA{p:lor-phi-94}, \citeA{p:and-gyh-08} and \citeA{p:berk-09}), or 
comparisons of wavelet coefficients (\citeA{p:sah-neu-99}, \citeA{p:cho-fry-09}). The underlying
important assumption, on which these tests are based, is on a delicate,
subjective, choice of segments of the data. In this paper we propose a
test based on the discrete Fourier transforms based on the entire length of data, thus
avoiding a subjective choice of segment length.

In Section \ref{sec:theteststatistic} 
we define the Discrete Fourier transform and show that the
Discrete Fourier transforms are asymptotically uncorrelated at the canonical
frequencies if and only if the time series is second order stationary. 
This motivates the test statistics, which is based on the discrete 
Fourier transform.  The Portmanteau type test statistics we propose
is based on the covariance function calculated using the 
discrete Fourier transform at the canonical frequencies. 
The asymptotic sampling distribution of the test statistic is
obtained in Section \ref{sec:properties-stat}. 
Further we show that the asymptotic sampling  distribution
is approximately distributed as a central chi-square under the null
hypothesis that the time series is second order stationary. To examine the
power of the test, we consider the case of locally stationary time series
(see \citeA{p:sub-70}, 
\citeA{p:oza-ton-75}, \citeA{b:pri-88}, \citeA{p:dah-97},  
\citeA{p:nas-99},  and \citeA{p:nas-00}), 
and derive the distribution of the test statistic
under this class of alternatives. In Section
\ref{sec:sampling-nonstat}
we show the distribution under this class of alternatives, is a type of 
non-central chi-square, where 
the noncentrality parameter is in some sense a measure of
departure of nonstationarity. In Section \ref{sec:sim}, 
through simulations, we examine
the power of the test, and show that for 
various types of alternatives the power is very high.
We end this section with various comments on the types of nonstationary
behaviour the test can detect. 
In Section \ref{sec:data} we illustrate
the test with two real data examples.

An outline of important aspects of some proofs can be found in the appendix, and the full
details can be found in the accompanying technical report.

\section{The Test Statistic, 
motivation and sampling distribution}\label{sec:theteststatistic}

\subsection{Motivation}

Let $\{X_{t}\}$ be
a zero mean time series.  Suppose we observe 
$\{X_{t};t=1,\ldots,T\}$ and let $J_{T}(\omega _{k})$ be
its Discrete Fourier Transform defined as 
\begin{eqnarray*}
J_{T}(\omega_{k}) = \frac{1}{\sqrt{2\pi T}}\sum_{t=1}^{T}X_{t}\exp(it\omega_{k}),
\textrm{ for } 1\leq k\leq T,
\end{eqnarray*}
where $\omega_{k} = \frac{2\pi k}{T}$ are the canonical frequencies. 
It is well known that if $\{X_{t}\}$ is a second order stationary time series, 
whose covariances 
are absolutely summable then for $k_{1}\neq k_{2}$ and $k_{1}\neq T-k_{2}$ 
we have 
$\cov\big(J_{T}(\omega_{k_{1}}), J_{T}(\omega_{k_{2}})\big) = O(\frac{1}{T}).$ 
Therefore in the case of stationary processes, the discrete Fourier transform
$\{J_{T}(\omega_{k})\}_{k=1}^{T}$
is asymptotically uncorrelated. Let 
\begin{eqnarray*}
\kappa(t,\tau) = \Ex(X_{t},X_{\tau}) = 
\frac{1}{T}\sum_{k_{1},k_{2}=1}^{T}\Ex(J_{T}(\omega_{k_{1}}) 
\overline{J_{T}(\omega_{k_{2}})})
\exp(-it\omega_{k_{1}}+i\tau\omega_{k_{2}}),
\end{eqnarray*}
where $\overline{z}$ is the complex conjugate of the complex variable $z$. 
From the above we observe
if $\Ex(J_{T}(\omega_{k_{1}})\overline{J_{T}(\omega_{k_{2}})})=0$ 
for $k_{1}\neq k_{2}$ or
$k_{1}\neq T-k_{2}$, then we have  
$\kappa(t,\tau)=\kappa(t-\tau)$ for $0\leq t,\tau \leq T-1$. 
In other words, an uncorrelated discrete Fourier transform sequence
implies that the original time series is
second order stationary, up to lag $T$.

Let us consider a simple example to show that even if the time series are
independent, but not stationary,  then $\{J_{T}(\omega_{k})\}$ are not
uncorrelated. 
Let us suppose $X_{t} = \sigma_{t}\varepsilon_{t}$, where $\sigma_{t}$ is a deterministic,  
time dependent function and $\{\varepsilon_{t}\}$ are independent
identically distributed random variables with $\Ex(\varepsilon_{t})=0$ and 
 $\var(\varepsilon_{t})=1$. 
In this case, the covariance of the discrete Fourier transform at the canonical frequencies is 
\begin{eqnarray*}
cov(J_{T}(\omega_{k_{1}}),J_{T}(\omega_{k_{2}})) = 
\frac{1}{2\pi T}\sum_{t=1}^{T}\sigma_{t}^{2}\exp(it(\omega_{k_{1}}-
\omega_{k_{2}})). 
\end{eqnarray*}
From the above, it is clear that 
$\cov(J_{T}(\omega_{k_{1}}),J_{T}(\omega_{k_{2}}))\neq 0$ for some 
$k_{1}\neq k_{2}$. If we suppose that
$\sigma_{t}$ is a  sample
from a smooth function $\sigma:[0,1]\rightarrow \mathbb{R}$, that is 
$\sigma_{t} = \sigma(t/T)$,  
then by using the rescaling methods often used in nonparametric statistics 
we have 
\begin{eqnarray*}
\cov(J_{T}(\omega_{k_{1}}),J_{T}(\omega_{k_{2}})) = 
\int_{0}^{1}\sigma(u)^{2}\exp(i2\pi u(k_{1}-k_{2}))du. 
\end{eqnarray*}

\subsection{The test statistic}

The above observations lead us to the following test statistic. We note
that, if the the time series is second order stationary, then 
$\Ex(J_{T}(\omega_{k})) = 0$ and $\var(J_{T}(\omega_{k})) \rightarrow 
f(\omega_{k})$ as $T\rightarrow \infty$, where 
$f:[0,2\pi]\rightarrow \mathbb{R}$ 
is the spectral density of the original time series $\{X_{t}\}$
(see \citeA{b:pri-88} and  \citeA{b:bro-dav-87}). 
Therefore by standardising with $\sqrt{f(\omega_{k})}$, under the 
null of stationarity, $\{J_{T}(\omega_{k})/\sqrt{f(\omega_{k})}\}$ is close to 
an uncorrelated, second order stationary sequence. 
Therefore to test for stationarity of 
$\{X_{t}\}$ we will test for randomness of the sequence 
$\{J_{T}(\omega_{k})/\sqrt{f(\omega_{k})}\}$. The proposed test will be a 
type of Portmanteau test (see \citeA{p:che-deo-04} for 
applications of the Portmanteau test
in time series analysis). 
Of course, in reality the spectral density
$f(\omega)$ is unknown, therefore we will replace $f(\cdot)$ with the 
estimated spectral density
function $\widehat{f}_{T}(\cdot)$, where 
\begin{eqnarray}
\label{eq:spectral-estimator}
\widehat{f}_{T}(\omega_{k}) = 
\sum_{j}\frac{1}{bT}K(\frac{\omega_{k}-\omega_{j}}{b})|J_{T}(\omega_{j})|^{2}, 
\end{eqnarray}
$K:[-1/2,1/2]\rightarrow \mathbb{R}$ 
is a positive, continuous, 
symmetric kernel function which satisfies $\int_{-1}^{1} K(x)dx=1$ and
$\int_{-1}^{1} K(x)^{2}dx <\infty$ and $b$ is a bandwidth.

We define the empirical covariance at lag $r$, which is complex valued, of the 
discrete Fourier transform as 
\begin{eqnarray}
\label{eq:cov}
\widehat{c}_{T}(r)=\frac{1}{T}\sum_{k=1}^{T}\frac{J_{T}(\omega _{k})
\overline{J_{T}(\omega _{k+r})}}{\sqrt{\widehat{f}_{T}(\omega _{k})\widehat{f}_{T}
(\omega _{k+r})}}, \quad \textrm{for } 1\leq r \leq T-1. 
\end{eqnarray}

The proposed test statistic is based on $\widehat{c}_{T}(r)$. In the technical report 
we show 
that if $\{X_{t}\}$ is a stationary time series and $\Ex(X_{t}^{4})<\infty$
then both the variance of the real and imaginary parts of
$\sqrt{T}\widehat{c}_{T}(r)$ converge to 
\begin{eqnarray}\label{eq:var-nonlinear}
1 +\frac{1}{4\pi}\int_{0}^{2\pi}
\int_{0}^{2\pi}\frac{f_{4}(\lambda_{1},-\lambda_{1}-\omega_{r},\lambda_{2})}{
\sqrt{f(\lambda_{1})f(\lambda_{1}+\omega_{r})f(\lambda_{2})f(\lambda_{2}+\omega_{r})}}d\lambda_{1}d\lambda_{2}, 
\end{eqnarray}
as $T\rightarrow \infty$, 
where $\omega_{r} = 2\pi r/T$, 
$f_{4}(\lambda_{1},\lambda_{2},\lambda_{3}) = (2\pi)^{-3}
\sum_{j_{1},j_{2},j_{3}=-\infty}^{\infty}\cum(X_{0},X_{j_{1}},X_{j_{2}},X_{j_{3}})
\exp(i(\lambda_{1}j_{1} + \lambda_{2}j_{2} + \lambda_{3}j_{3}))$ is the fourth order
cumulant spectra. 
Furthermore, under the null hypothesis of second order 
stationarity, we show in 
Lemma \ref{lemma:variance-stat}, that 
the empirical covariances $\widehat{c}_{T}(r)$ at different lags
are asymptotically uncorrelated and $\widehat{c}_{T}(r)=o_{p}(1)$. 
Therefore we define the test statistic
\begin{displaymath}
\label{eq:TeststatisticNon}
\mathcal{T}_{m} = 
T\sum_{n=1}^{m}
\frac{|\widehat{c}_{T}(r_{n})|^{2}}
{1 + \frac{1}{4\pi}\int_{0}^{2\pi} \int_{0}^{2\pi}
\frac{f_{4}(\lambda_{1},-\lambda_{1}-\omega_{r_{n}},-\lambda_{2})}{
\sqrt{f(\lambda_{1})f(\lambda_{1}+\omega_{r_{n}})f(\lambda_{2})
f(\lambda_{2}+\omega_{r_{n}})}}d\lambda_{1}d\lambda_{2}},
\end{displaymath}
where $|z|^{2} = z\overline{z}$ and $r_{n}\neq 0$ or $T/2$. 
For example, we can choose $r_{n}=n$, and use $m$ consecutive lags. 
We note, that unlike the classical Portmanteau tests, using 
covariances with a large lag is not problematic as the discrete Fourier transform is 
periodic. 

We derive the asymptotic distribution of the test statistic in Section
\ref{sec:properties-stat}, under the null hypothesis that 
$\{X_{t}\}$ statisfies the MA$(\infty)$ representation
\begin{eqnarray}
\label{eq:ma}
X_{t} = \sum_{j=0}^{\infty}\psi_{j}\varepsilon_{t-j}, 
\end{eqnarray}
where $\{\varepsilon_{t}\}$ are independent, identically distributed
random variables with $\Ex(\varepsilon_{t})=0$, $\Ex(\varepsilon_{t}^{2})=1$ and 
$\kappa_{4}=\cum_{4}(\varepsilon_{t})$. 
Under these assumptions we will show in Corollary \ref{cor:asymptotic-dis}, 
below, that $\widehat{c}_{T}(r) = o_{p}(1)$ and $\mathcal{T}_{m}$ converges in 
distribution to $\chi_{2m}^{2}$. 
Therefore we reject the null of second order stationarity at the 
$\alpha\%$ significance level if $\mathcal{T}_{m} > \chi^{2}_{2m}(1-\alpha)$. 
We proved the above result under the assumption that the time series 
is stationary, linear and has absolutely summable covariances. But we believe this 
result is true even if the process is nonlinear, but stationary or 
has long memory but is stationary. The proof is beyond the scope of 
this paper. We need strong mixing condition to prove this general result.

In the case of linearity, (\ref{eq:var-nonlinear}) has an interesting form. 
It can be shown that 
\begin{eqnarray*}
\frac{1}{2\pi}
\int_{0}^{2\pi}
\int_{0}^{2\pi}\frac{f_{4}(\lambda_{1},-\lambda_{1}-\omega_{r},\lambda_{2})}{
\sqrt{f(\lambda_{1})f(\lambda_{1}+\omega_{r})f(\lambda_{2})f(\lambda_{2}+\omega_{r})}}
d\lambda_{1}d\lambda_{2}
= \kappa_{4}
\bigg|\frac{1}{2\pi}\int_{0}^{2\pi}\exp(i\phi(\omega)-\phi(\omega+\omega_{r}))
d\omega\bigg|^{2},  
\end{eqnarray*}
where 
\begin{eqnarray*}\phi(\omega) = \arctan \frac{\Im A(\omega)}{\Re A(\omega)} \textrm{ and }
A(\omega) = \frac{1}{\sqrt{2\pi}}\sum_{j=0}^{\infty}\psi_{j}\exp(i \omega j).
\end{eqnarray*} 
Hence in the case of linearity, the test statistic is equivalent to 
\begin{eqnarray}
\label{eq:Teststatistic}
\mathcal{T}_{m}  = T\sum_{n=1}^{m}
\frac{|\widehat{c}_{T}(r_{n})|^{2}}{1+\frac{\kappa_{4}}{2}
\big|\frac{1}{2\pi}
\int_{0}^{2\pi}\exp(i\phi(\omega)-\phi(\omega+\frac{2\pi r_{n}}{T}))
d\omega\big|^{2}}. 
\end{eqnarray}
Morever, if $m$ is small and small lags are used (ie. $r_{n}=n$)
then the test statistic can be approximated by 
$\mathcal{T}_{m}  = T\sum_{r=1}^{m}|\widehat{c}_{T}(r)|^{2}/(1+\kappa_{4}/2)$. 


\begin{remark}[Estimation of the tri-spectra]\label{remark:parameter}
We observe that the test statistic $\mathcal{T}_{m}$ requires  estimates of the  
the parameter
\begin{eqnarray*}
\kappa_{r} = \frac{1}{2\pi}
\int_{0}^{2\pi} \int_{0}^{2\pi}\frac{f_{4}(\lambda_{1},-\lambda_{1}-\omega_{r},-\lambda_{2})}{
\sqrt{f(\lambda_{1})f(\lambda_{1}+\omega_{r})
f(\lambda_{2})f(\lambda_{2}+\omega_{r})}}d\lambda_{1}d\lambda_{2}. 
\end{eqnarray*}
Therefore to estimate the above parameter we require estimators of the 
tri-spectra and spectral density $f_{4}$ and $f$ respectively. 
\citeA{p:bri-ros-67} propose a consistent estimator of the tri-spectra $f_{4}(\cdot)$,
which we denote as $\widehat{f}_{4,T}(\cdot)$. Therefore an estimator of the above 
integral is
\begin{eqnarray*}
\hat{\kappa}_{r,T} = \frac{1}{2\pi}
\int_{0}^{2\pi} \int_{0}^{2\pi}\frac{\widehat{f}_{4,T}(\lambda_{1},-\lambda_{1}-\omega_{r},-\lambda_{2})}{
\sqrt{\widehat{f}_{T}(\lambda_{1})\widehat{f}_{T}(\lambda_{1}+\omega_{r})
\widehat{f}_{T}(\lambda_{2})\widehat{f}_{T}(\lambda_{2}+\omega_{r})}}d\lambda_{1}d\lambda_{2}, 
\end{eqnarray*}
where $\widehat{f}_{T}(\lambda_{2})$ is defined in
(\ref{eq:spectral-estimator}). Since $\hat{\kappa}_{r,T}$ is a consistent estimator of 
$\kappa_{r}$, replacing  $\kappa_{r}$ in the test statistic with $\hat{\kappa}_{r,T}$, does not 
alter the asymptotic sampling distribution.  
\end{remark}

\begin{remark}[Practical issues]
The asymptotic distribution under the null is derived under the assumptions that the 
spectral density of the time series $\{X_{t}\}$ is bounded away from zero. In practice, 
even if this assumption holds, the estimated spectral density $\widehat{f}_{T}(\cdot)$
may be quite close to zero. Therefore, in this case, 
to prevent falsely rejecting the null, we suggest adding a small ridge to the spectral 
density estimator $\widehat{f}_{T}(\cdot)$
to bound it away from zero. 
\end{remark}

\subsection{The power of the test}

In Section \ref{sec:sampling-nonstat} we obtain the asymptotic sampling
properties of the test statistic $\mathcal{T}_{m}$, under the alternative 
of local stationarity. In order to understand what nonstationary behaviour 
the test statistic can detect and how to select the lag $r$ in the 
test statistic, we will now outline some of the results in 
 Section \ref{sec:sampling-nonstat}.
Suppose that $\{X_{t}\}$ is a nonstationary time series, 
where in a small neighbourhood of $t$ the observations are close to stationary 
and has the local spectral density $f(\frac{t}{T},\omega)$.
In Lemma \ref{lemma:meanalt} we show that $\widehat{c}_{T}(r)\approx B(r)$, 
where
\begin{eqnarray}
\label{eq:bias-LS}
B(r) = 
\frac{1}{2\pi}\int_{0}^{2\pi} \int_{0}^{1}
\frac{1}{[f(\lambda)f(\lambda + \omega_{r})]^{1/2}}
f(u,\lambda)\exp(-i\frac{2\pi t}{T} r)du d\lambda,
\end{eqnarray}
and that $\mathcal{T}_{m}$ has asymptotically a type of non-central chi squared distribution
where the noncentrality parameters are given by $\{B(r_{n})\}$. 
Hence, the test statistic is more likely to reject the null, the further 
$B(r_{n})$, is from from zero. Studying the above we see that  
if the dynamics change slowly over time, then a small lag $r_{n}$, should yield 
a large $B(r)$. On the other hand,  if there is a rapid
change in the behaviour, a large $r_{n}$, leads to 
a large $B(r_{n})$. Therefore, in this case, by
using a large $r_{n}$, we are more likely to reject the null. 


\section{Sampling properties of the test statistic under the null}\label{sec:properties-stat}

We now derive the asymptotic distribution under the null of stationarity. 
We will use the following assumptions. 

\begin{assumption}\label{assum:stat}
Let us suppose that $\{X_{t}\}$ satisfies (\ref{eq:ma}).

Let $A(\omega) = (2\pi)^{1/2}
\sum_{j=0}^{\infty}\psi_{j}\exp(ij\omega)$ and define the spectral density
$f(\omega) = |A(\omega)|^{2}$. 
Assume 
\begin{itemize}
\item[(i)] $\sum_{j=0}^{\infty}|j\psi_{j}|<\infty$ (noting that this
implies $\sup_{\omega}|f^{\prime}(\omega)| < \infty$). 
\item[(ii)] $\Ex(|\varepsilon_{t}^{16}|)<\infty$. 
\item[(iii)] $\inf_{\omega} f(\omega) > 0$ and $\inf_{\omega} |\Re A(\omega)|^{2} > 0$.  
\item[(iv)] Either (a) $\sum_{j}|j^{2}\psi_{j}|<\infty$ or  (b) 
the derivative of the spectrum $f^{\prime}(\omega)$
is piecewise montone on the interval $[0,2\pi]$ (in other words $f^{\prime}(\cdot)$ can be 
partitioned into a finite number of intervals which is either nonincreasing or 
nondecreasing).  
\end{itemize}
\end{assumption}

We use Assumption \ref{assum:stat}(i,ii) to show asymptotic normality if 
$\widehat{c}_{T}(r)$ (in fact Assumption \ref{assum:stat}(ii) is used to 
obtain the error when replacing $\widehat{c}_{T}(r)$, with $\widetilde{c}_{T}(r)$, defined
below). We use Assumption \ref{assum:stat}(iv) to obtain the rate of decay of the 
Fourier coefficients of the function $\frac{1}{\sqrt{f(\omega)f(\omega+\omega_{r})}}$. 

To simplify the analysis of the test statistic $\mathcal{T}_{m}$ we replace the denominator
in the covariance $\widehat{c}_{T}(r)$ with its deterministic limit. 
To do this, we 
define the unobserved covariance 
\begin{eqnarray}
\label{eq:cov-unob}
\widetilde{c}_{T}(r)=\frac{1}{T}\sum_{k=1}^{T}\frac{J_{T}(\omega_{k})
\overline{J_{T}(\omega_{k+r})}}{\sqrt{f(\omega_{k})f(\omega_{k+r})}},
\end{eqnarray}
and obtain the asymptotic distribution of $\widetilde{c}_{T}(r)$. 

In the following lemma we show that the difference between 
$\sqrt{T}|\widehat{c}_{T}(r) -  \widetilde{c}_{T}(r)|$ is negligible. 
\begin{theorem}\label{lemma:cov-diff}
Suppose Assumption \ref{assum:stat} is statisfied and let  $\widehat{c}_{T}(r)$ and 
$\widetilde{c}_{T}(r)$ be defined as in (\ref{eq:cov}) and (\ref{eq:cov-unob}) respectively. 
Then we have 
\begin{eqnarray}
\label{eq:thm3.1}
\sqrt{T}|\widehat{c}_{T}(r) -  \widetilde{c}_{T}(r)| =
O_{p}\bigg((b+\frac{1}{\sqrt{bT}}) + \big(\frac{1}{bT^{1/2}} + b^{2}T^{1/2}\big)
\big(\frac{1}{|r|} + \frac{1}{T^{1/2}}\big) \bigg). 
\end{eqnarray}
\end{theorem}
PROOF. See Appendix \ref{sec:proof-diff}, Lemma \ref{lemma:diff-both}, equation
(\ref{eq:thm3.1a}).  \hfill $\Box$

\vspace{3mm}


\vspace{3mm}

In the following lemma we derive the asymptotic variance of 
$\tilde{c}_{T}(r)$, and show that $\tilde{c}_{T}(r)$ is 
asymptotically uncorrelated at different lags $r$, and at the
real and imaginary parts. 

\begin{lemma}\label{lemma:variance-stat}
Suppose Assumption \ref{assum:stat} holds. Then we have 
\begin{eqnarray*}
&& \cov(\sqrt{T}\Re \tilde{c}_{T}(r_{1}),\sqrt{T}\Re \tilde{c}_{T}(r_{2})) = 
\cov(\sqrt{T}\Im \tilde{c}_{T}(r_{1}),\sqrt{T}\Im \tilde{c}_{T}(r_{2})) \\
&=& \left\{
\begin{array}{cc}
 O(\frac{1}{T}) & r_{1}\neq r_{2} \\
1 + \frac{\kappa_{4}}{2}
\bigg|\frac{1}{2\pi}
\int_{0}^{2\pi}\exp(i\phi(\omega)-\phi(\omega+\frac{2\pi r_{1}}{T}))
d\omega\bigg|^{2} + O(\frac{1}{T}) & r_{1} = r_{2} \\
\end{array}
\right.,
\end{eqnarray*}
and for all $r_{1},r_{2}\in \mathbb{Z}$, 
$\cov(\sqrt{T}\Im \tilde{c}_{T}(r_{1}),\sqrt{T}\Re \tilde{c}_{T}(r_{2})) = O(\frac{1}{T})$.
\end{lemma}
PROOF. See Appendix \ref{sec:var}. \hfill $\Box$

\vspace{3mm}

We now show normality of $\widehat{c}_{T}(r)$, which we use to obtain 
the distribution of $\mathcal{T}_{m}$.  
\begin{theorem}\label{thm:cCLT}
Suppose Assumption \ref{assum:stat} holds. Then for fixed $m$ we have 
\begin{eqnarray}
\label{eq:cCLTb}
\sqrt{T}\bigg(\frac{1}{1+\frac{\kappa_{4}}{2}\varphi(\frac{r_{1}}{T})}
\Re \widehat{c}_{T}(r_{1}),\ldots,
\frac{1}{1+\frac{\kappa_{4}}{2}\varphi(\frac{r_{m}}{T})}\Im \widehat{c}_{T}(r_{m})\bigg)
\Dcon \mathcal{N}(0,I_{2m}),
\end{eqnarray}
as 
$m(b+\frac{1}{\sqrt{bT}}) + \big(\frac{1}{bT^{1/2}} + b^{2}T^{1/2}\big)
\sum_{r=1}^{m}\big(\frac{1}{|r_{m}|} + \frac{1}{T^{1/2}}\big)\rightarrow 0$ and 
$T\rightarrow \infty$,
where $I_{2m}$ is the identity matrix and 
$\varphi(x) = 
\big|\frac{1}{2\pi}
\int_{0}^{2\pi}\exp(i\phi(\omega)-\phi(\omega+x))d\omega\big|^{2}$.
\end{theorem}
PROOF. See Appendix \ref{sec:asym-normality}. \hfill $\Box$

\vspace{3mm}
By using the above we are able to obtain the asymptotic distribution of $\mathcal{T}_{m}$.
\begin{corollary}\label{cor:asymptotic-dis}
Suppose Assumption \ref{assum:stat} holds. Then for fixed $m$ we have 
$\mathcal{T}_{m} \Dcon \chi_{2m}^{2}$ 
with
\\* 
$m(b+\frac{1}{\sqrt{bT}}) + \big(\frac{1}{bT^{1/2}} + b^{2}T^{1/2}\big)
\sum_{n=1}^{m}\big(\frac{1}{|r_{n}|} + \frac{1}{T^{1/2}}\big)\rightarrow 0$,
as $T\rightarrow \infty$.
\end{corollary}
PROOF. The result immediately follows from Theorem \ref{thm:cCLT}. \hfill $\Box$

\vspace{3mm}
Hence we have shown under the null, that the test statistics has asymptotically
a $\chi^{2}$ distribution. 

\section{Sampling properties of the test statistic under the alternative of 
local stationarity}\label{sec:sampling-nonstat}

It is useful to investigate the behaviour of the test statistic
in the case that the null does not hold. If the covariance structure
varies over time, then the limit of $\widehat{c}_{T}(r)$ will not 
be zero. This suggests that the test statistic will have a type of 
non-central $\chi^{2}$ distribution. 
However, in the case that time-varying covariance has no 
structure it is not clear what the limit of $\widehat{c}_{T}(r)$ will be. 
For example, consider the 
simple example of a time-varying AR process,  
$X_{t} = a(t)X_{t-1} + \varepsilon_{t}$, where
$\{\varepsilon_{t}\}$ are iid random variables. Without any structure on the 
AR coefficient, $a(t)$, it is not clear what the spectral density estimator, 
$\widehat{f}_{T}(\cdot)$, defined in (\ref{eq:spectral-estimator}) 
should converge to. Hence it is not possible to obtain the limit of
$\widehat{c}_{T}(r)$. On the other hand, let us suppose that $a(t)$, varies
slowly over time and $a(t)$ is a sample from a function $a:[0,1]\rightarrow
\mathbb{R}$, that is for some $T$, $a(t) = a(\frac{t}{T})$, and the time series
satisfies
$X_{t,T} = a(\frac{t}{T})X_{t-1,T} + \varepsilon_{t}, \quad 
t=1,\ldots, T$.  
Now in this set up, as we let $T\rightarrow \infty$, $a(\cdot)$ varies 
less and $X_{t,T}$ is observed on a finer grid, which in reality can never
be realised. Hence, by supposing that $a(\cdot)$ varies slowly over time
we have imposed some structure on the time-varying parameter and 
we are using an infill asymptotic argument (see \citeA{p:rob-89}). 
In this case, we will show, below, that 
$\widehat{f}_{T}(\cdot)$ is an estimator of the integrated spectrum,
this can also be viewed as the average of local spectrums. 
The model described above, 
is an example of a locally stationary linear time series considered in 
\citeA{p:dah-97} and  \citeA{p:dah-pol-06}. 
Therefore, following \citeA{p:dah-pol-06}, 
we define a locally stationary linear time series as 
\begin{eqnarray}
\label{eq:tvma}
X_{t,T} = \sum_{j=0}^{\infty}\psi_{t,T}(j)\varepsilon_{t-j},
\end{eqnarray} 
where $\{\varepsilon_{t}\}$ are iid random variables, 
$\Ex(\varepsilon_{t})=0$, $\var(\varepsilon_{t})=1$. Therefore 
we will consider the behaviour of the test statistic under the alternative of 
local stationarity. 

In order for $\{X_{t,T}\}$ to be a locally stationary time series,
we will assume that $\psi_{t,T}(j)$ closely approximates 
the smooth function $\psi_{j}(\cdot)$. 
Hence the time-varying MA parameters $\{\psi_{t,T}(j)\}$ vary slowly over time. 
It can be shown that in this case, 
$\{X_{t,T}\}$ is a locally stationary time series because
it can locally be approximed by a stationary time series.
We will use the following assumptions. 

\begin{assumption}\label{assum:nonstat}
Let us suppose that $\{X_{t,T}\}$ satisfies (\ref{eq:tvma}). 
Suppose, there exists a 
sequence of functions $\psi_{j}(u)$, such that $\psi_{j}(u)$ is Lipschitz continuous and 
$|\psi_{j}(\frac{t}{T}) - \psi_{t,T}(j)|\leq T^{-1}\ell(j)^{-1}$, where 
$\{\ell(j)^{-1}\}$ is a positive monotonically decreasing function which 
satisfies $\sum_{j}|j|^{2}\ell(j)^{-1} < \infty$. Let  
$f(u,\omega) = (2\pi)^{-1}|\sum_{j=0}^{\infty}\psi_{j}(u)\exp(ij\omega)|^{2}$ (hence 
$\sup_{u,\omega}|\frac{\partial^{2} f(u,\omega)}{\partial \omega^{2}}|<\infty$). 
\begin{itemize}
\item[(i)] $\sup_{u}|\psi_{j}(u)|<\ell(j)^{-1}$ and 
$\sup_{u}\big|\frac{d \psi_{j}(u)}{du}\big| < K\ell(j)^{-1}$ (hence 
$\sup_{u,\omega}|\frac{\partial f(u,\omega)}{\partial u}|<\infty$). 
\item[(ii)] $\Ex(|\varepsilon_{t}^{16}|)<\infty$. 
\item[(iii)] Define the integrated spectral density
$f(\omega) = \int_{0}^{1}f(u,\omega)du$, and assume that 
$\inf_{\omega} f(\omega) > 0$. 
\item[(iv)] Either (a) $\sup_{u}\sum_{j}|\psi_{j}^{\prime\prime}(u)| < 
\infty$ (hence $\sup_{u,\omega}|\frac{\partial^{2} 
f(u,\omega)}{\partial u^{2}}|<\infty$) or 
(b) $A(u,\omega)$ and $\frac{\partial A(u,\omega)}{\partial u}$
are piecewise monotone functions with respect to $u$. 
\end{itemize}
\end{assumption}

We will show in Lemma \ref{lemma:meanalt}, below, 
that in the locally stationary case 
the spectral density estimator $\widehat{f}_{T}(\cdot)$ defined in 
(\ref{eq:spectral-estimator}) estimates the integrated spectrum $f(\omega)$, 
where 
$f(\omega)$ is defined in Assumption \ref{assum:nonstat}(iii). Roughly speaking, 
one can consider the integrated spectrum as the average of the locally stationary
spectrums. 

As in Section \ref{sec:properties-stat}, it is difficult to directly obtain the
distribution of $\widehat{c}_{T}(r)$. Instead we replace 
the random 
denominator with its deterministic limit (that is 
$J_{T}(\omega_{k})/\sqrt{\widehat{f}_{T}(\omega_{k})}$ with 
$J_{T}(\omega_{k})/\sqrt{f(\omega_{k})}$), and define 
\begin{eqnarray}
\label{eq:cov-unob1}
\widetilde{c}_{T}(r)=\frac{1}{T}\sum_{k=1}^{T}\frac{J_{T}(\omega_{k})
\overline{J_{T}(\omega_{k+r})}}{\sqrt{f(\omega_{k})f(\omega_{k+r})}}, 
\end{eqnarray}
where $f(\cdot)$ is the integrated spectrum. 
The following result is the locally stationary analogue of 
Theorem \ref{lemma:cov-diff}.

\begin{theorem}\label{lemma:cov-diff-nonstat}
Suppose Assumption \ref{assum:stat} is statisfied, and let  $\widehat{c}_{T}(r)$ and 
$\widetilde{c}_{T}(r)$ be defined as in (\ref{eq:cov}) and (\ref{eq:cov-unob1}) respectively. 
Then we have 
\begin{eqnarray*}
\sqrt{T}|\widehat{c}_{T}(r) -  \widetilde{c}_{T}(r)| =
O_{p}\bigg(\frac{1}{\sqrt{bT}} + \big(\frac{1}{bT^{1/2}} + b^{2}T^{1/2}\big)
\big(\frac{1}{|r|} + \frac{1}{T^{1/2}}\big)
\bigg). 
\end{eqnarray*}
\end{theorem}
PROOF. In Appendix \ref{sec:proof-diff}, 
Lemma \ref{lemma:diff-both}, equation (\ref{eq:thm4.1b}). \hfill $\Box$

\vspace{3mm}
From the lemma above we see that in order for the sampling properties of 
$\sqrt{T}\widehat{c}_{T}(r)$ and $\sqrt{T}\widetilde{c}_{T}(r)$ to coincide, 
we require that $T^{-1/2} << b << T^{-1/4}$.

We now obtain the mean and variance of  $\tilde{c}_{T}(r)$ under 
the alternative hypothesis of local stationarity. 
\begin{lemma}\label{lemma:meanalt}
Suppose Assumption \ref{assum:nonstat} are satisfied and 
let $f(\omega)$ and $f(u,\omega)$ be the integrated and local spectrum
(defined in  Assumption \ref{assum:nonstat}) respectively. 
Then we have 
\begin{eqnarray}
\label{eq:integrated-spec}
\Ex\big( \hat{f}_{T}(\omega) - f(\omega) \big)^{2} = O\big(b^{2} + \frac{1}{bT}\big), 
\end{eqnarray}
\begin{eqnarray}
\label{eq:bias-LS}
\Ex(\tilde{c}_{T}(r)) &\rightarrow& \frac{1}{2\pi}\int_{0}^{2\pi} 
\frac{1}{[f(\omega)f(\omega + \omega_{r})]^{1/2}}\bigg(
\int_{0}^{1} f(u,\omega)\exp(-2i\pi ru)du\bigg)d\omega, 
\end{eqnarray}
as $T\rightarrow \infty$, and 
\begin{eqnarray}
\cov(\Re\sqrt{T}\tilde{c}_{T}(r_{1}),\Re\sqrt{T}\tilde{c}_{T}(r_{2})) &\rightarrow& 
\Sigma^{(1,1)}_{T,r_{1},r_{2}} \quad 
\cov(\Re\sqrt{T}\tilde{c}_{T}(r_{1}),\Im\sqrt{T}\tilde{c}_{T}(r_{2})) \rightarrow 
\Sigma^{(1,2)}_{T,r_{1},r_{2}} \nonumber\\
\cov(\Im\sqrt{T}\tilde{c}_{T}(r_{1}),\Im\sqrt{T}\tilde{c}_{T}(r_{2})) &\rightarrow &
\Sigma^{(2,2)}_{T,r_{1},r_{2}}, \label{eq:limitvar-nonstat}
\end{eqnarray}
$b\rightarrow 0$, $bT\rightarrow \infty$ as  $T\rightarrow \infty$, where 
\begin{eqnarray*}
\Sigma^{(1,1)}_{T,r_{1},r_{2}} = 
\frac{1}{4}\big(\Gamma_{T,r_{1},r_{2}}^{(1)} + \Gamma_{T,r_{1},r_{2}}^{(2)}
+ \Gamma_{T,r_{2},r_{1}}^{(2)} + \Gamma_{T,r_{1},r_{2}}^{(3)} \big) +
O(\frac{\log T}{T}),
\end{eqnarray*}
\begin{eqnarray*}
\Sigma^{(1,2)}_{T,r_{1},r_{2}} = 
\frac{-i}{4}\big(\Gamma_{T,r_{1},r_{2}}^{(1)} + \Gamma_{T,r_{1},r_{2}}^{(2)}
- \Gamma_{T,r_{2},r_{1}}^{(2)} - \Gamma_{T,r_{1},r_{2}}^{(3)} \big) + O(\frac{\log T}{T}),
\end{eqnarray*}
\begin{eqnarray*}
\Sigma^{(2,2)}_{T,r_{1},r_{2}} = 
\frac{1}{4}\big(\Gamma_{T,r_{1},r_{2}}^{(1)} - \Gamma_{T,r_{1},r_{2}}^{(2)}
- \Gamma_{T,r_{2},r_{1}}^{(2)} + \Gamma_{T,r_{1},r_{2}}^{(3)} \big) + O(\frac{\log T}{T}),
\end{eqnarray*}
and  $\Gamma_{T,r_{2},r_{1}}^{(i)}$ ($i=1,2,3$) are defined in Lemma \ref{lemma:local-stat-cov}
(in Appendix \ref{sec:var}).  
\end{lemma}
PROOF. See Appendix \ref{sec:var}. \hfill $\Box$

\vspace{3mm}
We use the above to obtain the asymptotic distribution of $\mathcal{T}_{m}$
under the alternative. First we recall that we estimated the standardisation of 
$\widehat{c}_{T}(r)$, $\kappa_{r}$, in Remark \ref{remark:parameter}. It is worth noting that 
in the case that of local stationarity, $\hat{\kappa}_{r,T}$ is an estimator of 
\begin{eqnarray*}
\kappa_{r} = \frac{1}{2\pi}
\int_{0}^{2\pi} \int_{0}^{2\pi}
\frac{f_{4}(\lambda_{1},-\lambda_{1}-\omega_{r},-\lambda_{2})}{
\sqrt{f(\lambda_{1})f(\lambda_{1}+\lambda_{2})f(\lambda_{2})f(\lambda_{2}+\omega_{r})}}d\lambda_{1}d\lambda_{2}, 
\end{eqnarray*}
where $f(\cdot)$ is the integrated spectral density and 
and $f_{4}(\lambda_{1},\lambda_{2},\lambda_{3}) = \int_{0}^{1}
f_{4}(u,\lambda_{1},\lambda_{2},\lambda_{3})du$, with
$f_{4}(u,\lambda_{1},\lambda_{2},\lambda_{3}) = 
\frac{1}{2\pi}A(u,-\lambda_{1}-\lambda_{2}-\lambda_{3})
\prod_{i=1}^{3}A(u,\omega_{i})$. 

\begin{theorem}\label{thm:nonstatclt}
Suppose Assumption \ref{assum:nonstat} holds. Let 
\begin{eqnarray*}
\Sigma = 
\left(
\begin{array}{cc}
\Sigma^{(1,1)}_{T} & \Sigma^{(1,2)}_{T} \\
\Sigma^{(2,1)}_{T} & \Sigma^{(2,2)}_{T} \\
\end{array}
\right),
\end{eqnarray*}
where $\Sigma^{(1,1)}_{T,r_{1},r_{2}}$, $\Sigma^{(1,2)}_{T,r_{1},r_{2}}$, $\Sigma^{(2,2)}_{T,r_{1},r_{2}}$
are defined in Lemma \ref{lemma:meanalt} and 
$\Sigma^{(1,2)}_{T,r_{1},r_{2}} = 
\overline{\Sigma^{(2,1)}_{T,r_{1},r_{2}}}$. 

Furthermore define 
${\boldsymbol \mu}^{\prime} = (\Re B(r_{1}),\ldots,\Re B(r_{n}),\Im B(r_{1}),\ldots,\Im B(r_{n}))$, 
where 
\begin{eqnarray*}
B(r_{n}) = 
\frac{1}{2\pi}\int_{0}^{2\pi} \frac{1}{[f(\omega)f(\omega + \omega_{r_{n}})]^{1/2}}
\int_{0}^{1} f(u,\omega)\exp(-2i\pi r_{n}u)dud\omega. 
\end{eqnarray*}
Then we have 
\begin{eqnarray*}
\mathcal{T}_{m}\Dcon \sum_{n=1}^{m}\frac{\big(X_{n}^{2} + Y_{n}^{2}\big)}{
(1+\frac{1}{2}\kappa_{r_{n}})},
\end{eqnarray*}
with $\frac{m}{\sqrt{bT}} + \big(\frac{1}{bT^{1/2}} + b^{2}T^{1/2}\big)
\sum_{n=1}^{m}\big(\frac{1}{|r_{n}|} + \frac{1}{T^{1/2}}\big)\rightarrow 0$ as $T\rightarrow 0$, 
where ${\boldsymbol  X}_{2m}$ is a normally distributed random vector with
${\boldsymbol  X}_{2m} = (X_{1},\ldots,X_{m},Y_{1},\ldots, X_{m})^{\prime}$ 
and ${\boldsymbol X}_{2m} \sim \mathcal{N}({\boldsymbol \mu}, \Sigma)$.
Note the small abuse of notation, when we say $A\Dcon B$, we mean that 
the distribution of 
random variable $A$ converges to the distribution of random variable
$B$.  
\end{theorem}

\begin{remark}
We observe that if the matrix $\Sigma$, define in the 
theorem above, were the identity matrix, then the limiting distribution of 
$\mathcal{T}_{m}$, is a non-central $\chi^{2}_{2m}$ 
where the noncentrality parameter
is determined by the limit of $\widehat{c}_{T}(r_{n})$ (for $n=1,\ldots,m$). 
Hence the power of the  
test depends on the deviation of 
\\*
$\frac{1}{2\pi}\int_{0}^{2\pi} 
\frac{1}{[f(\omega)f(\omega + \omega_{r_{n}})]^{1/2}}
\int_{0}^{1} f(u,\omega)\exp(-2i\pi ru)dud\omega$ from zero, 
for each of the lags $r_{n}$. We see that term depends on the Fourier 
coefficient 
$a(r_{n};\omega) := \int_{0}^{1} f(u,\omega)\exp(-2i\pi r_{n}u)dud\omega$ and 
the magnitude of $a(r_{n};\omega)$ depends on whether the frequency of the  
nonstationary variation matches $r_{n}$. 

However, in the case that $X_{t,T}$ is s second order nonstationary 
time series, $\Sigma$ will not be a diagonal time series, because there
is correlation between the real and imaginary parts of $\widetilde{c}_{T}(r)$
and also correlation at different lags $r$. 
Thus the limiting distribution of
$\mathcal{T}_{m}$ will not be a classical non-central $\chi_{2m}^{2}$, 
due to the correlations in $\Sigma$. 
However, the conclusions discussed above still hold, namely the power of the test 
is determined, mainly, by the magnitude of 
mean vector ${\boldsymbol \mu}$. 
\end{remark}

\section{Simulations}\label{sec:sim}

We now consider a simulation study. We compare the results of 
the test statistic for both stationary and nonstationary time series. 
In each case, we replicate the time series
1000 times, and for each replication we do the test. 
We do the test for sample sizes $T=256$ and $T=512$. We do the test for 
$m =1 , 5$ and $10$. 
The percentage of time the test statistic exceeds
$\chi_{2m}^{2}(0.05)$, that is $\mathcal{T}_{m} > \chi_{2m}^{2}(0.05)$
is given in the tables, we also give plots of the empirical density of the 
test statistic. 

\subsection{Stationary time series}

\begin{figure}[h!]
\begin{center}
\includegraphics[width=8cm, height=5cm]{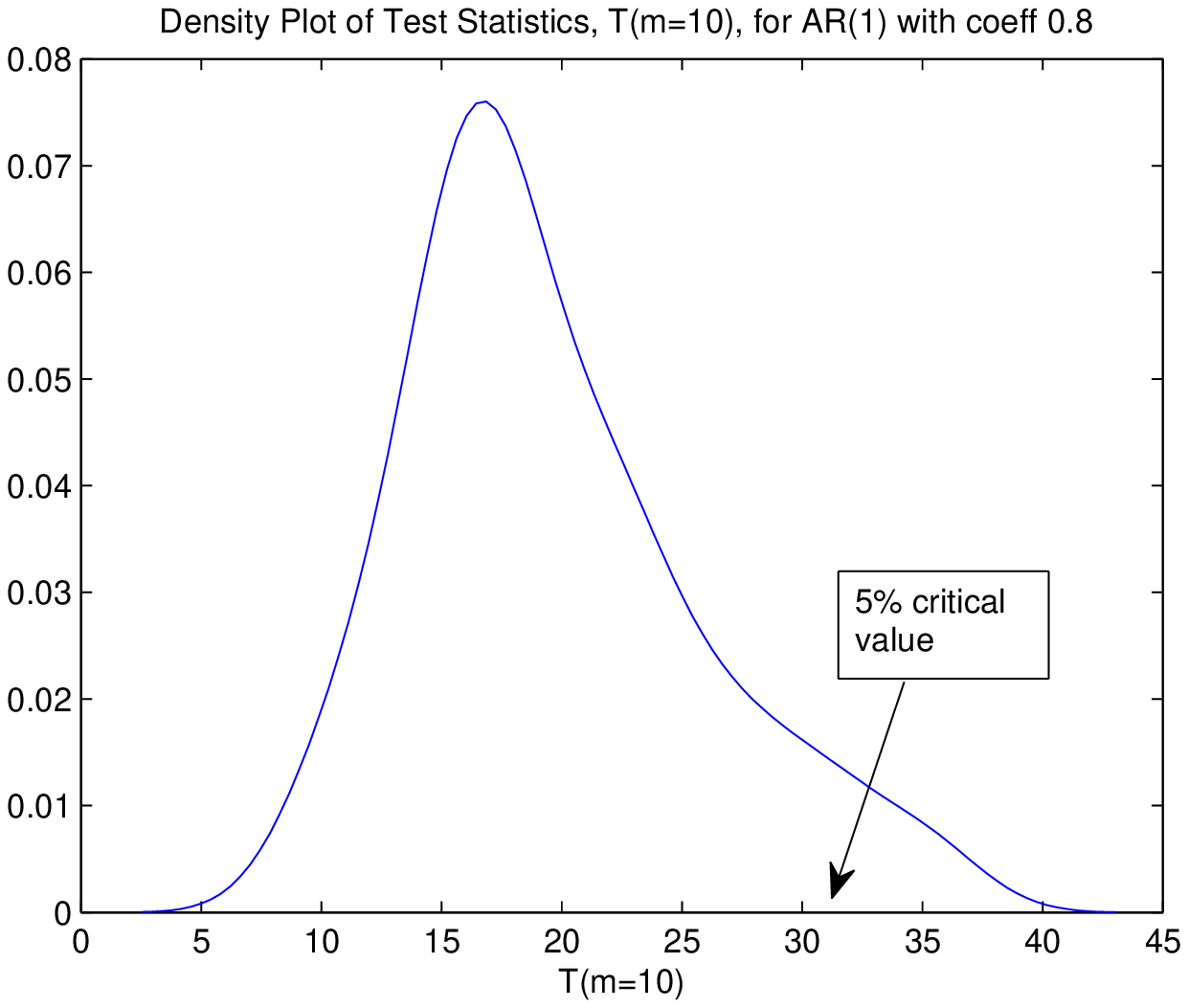}
\includegraphics[width=8cm, height=5cm]{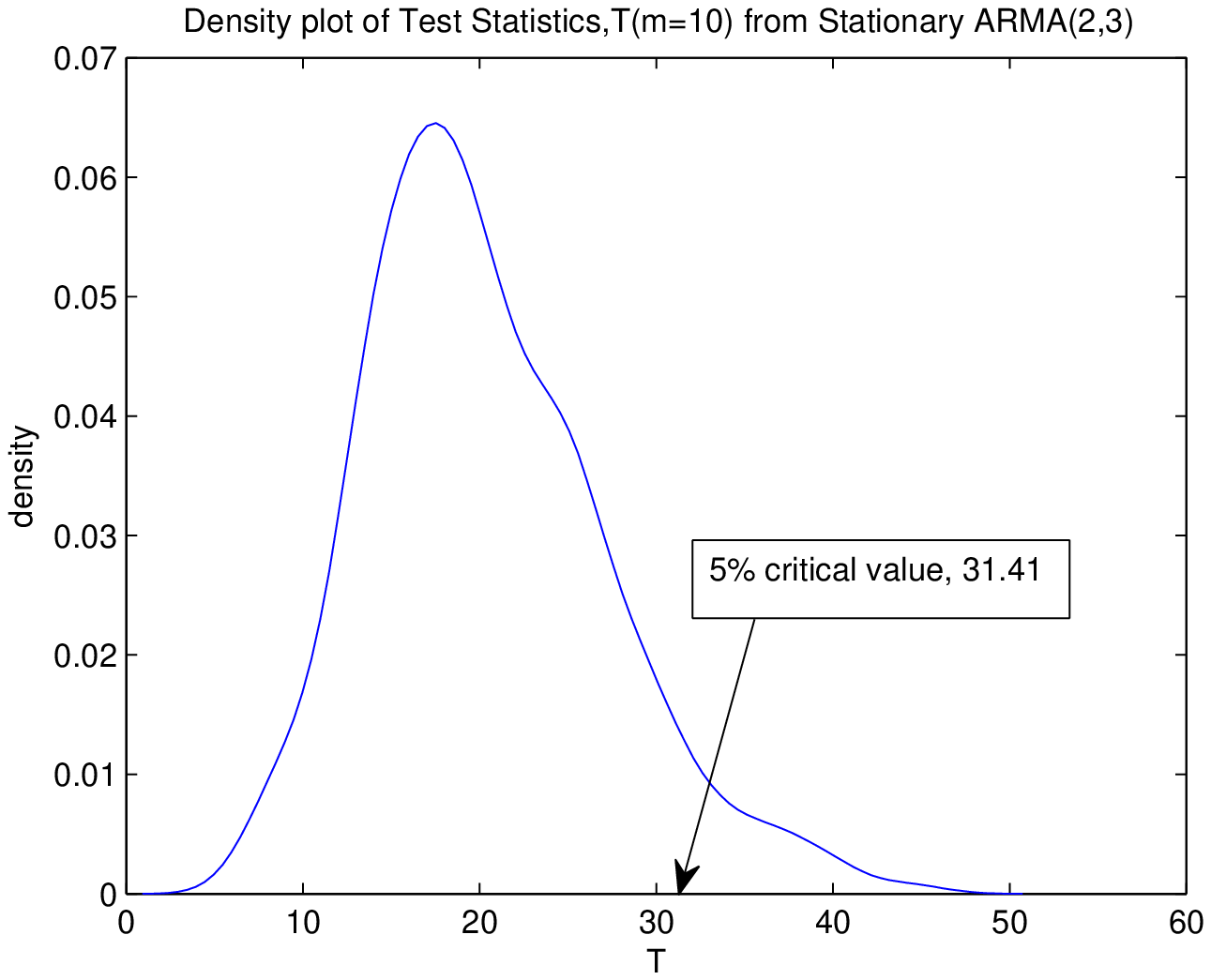}
\end{center}
\caption{\label{fig:1} The estimate density of the test statistic $\mathcal{T}_{10}$ for Left: 
Stationary Model 1 and Right: stationary Model 2 ($T=512$)}
\begin{center}
\includegraphics[width=8cm, height=5cm]{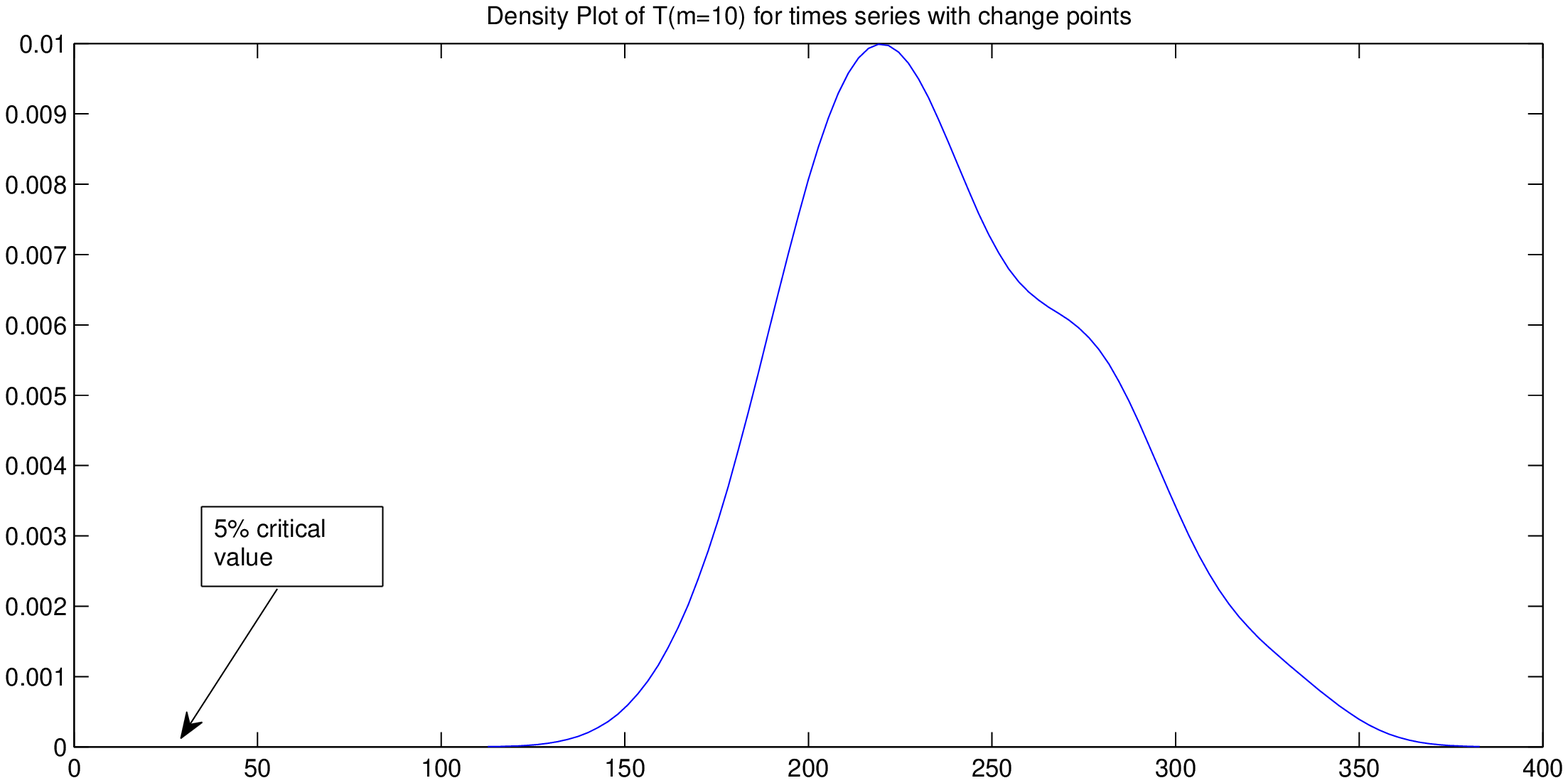}
\includegraphics[width=8cm, height=5cm]{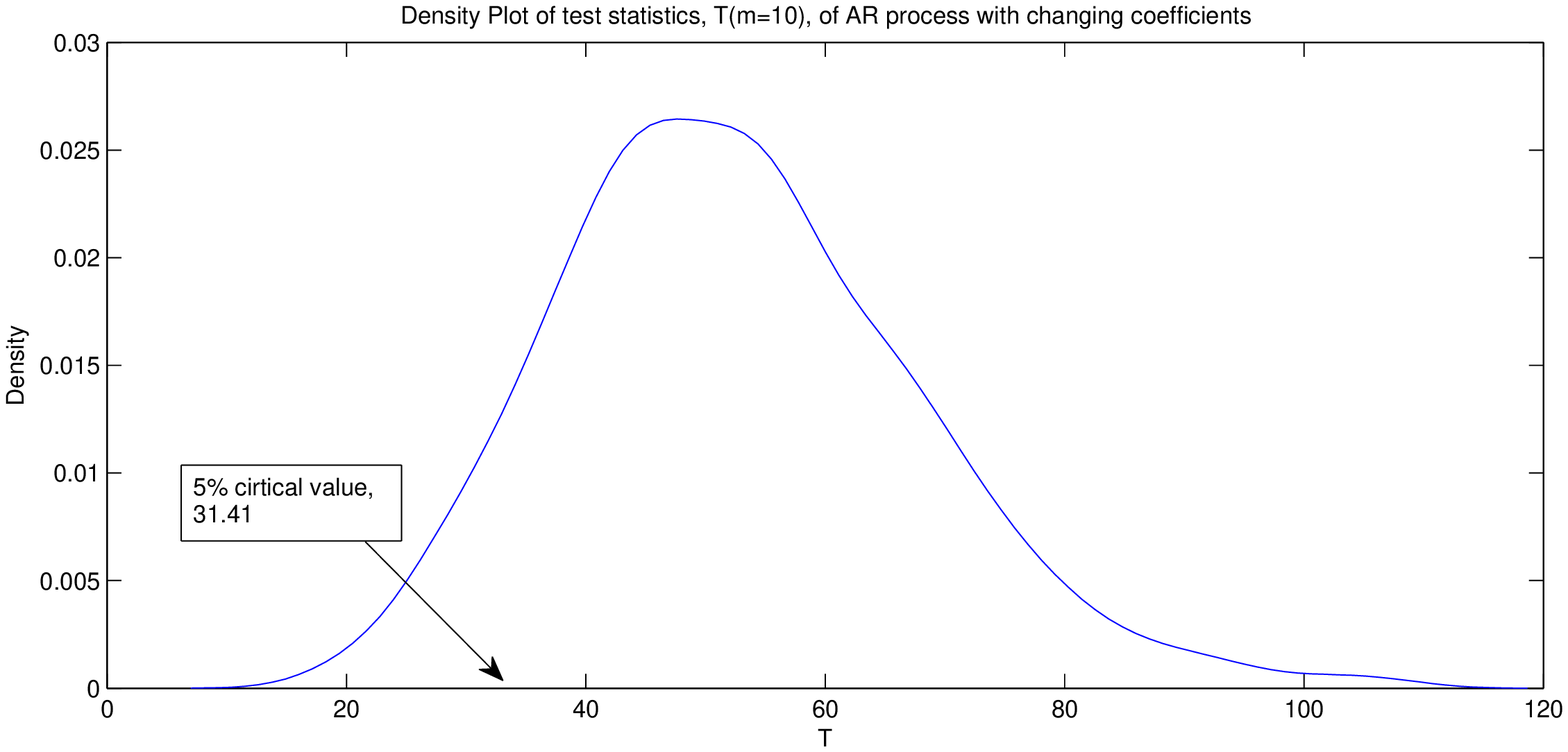}
\end{center}
\caption{\label{fig:2} Estimate density of test statistic for Left: 
Nonstationary Model 3 and Right: Nonstationary Model 4
(right). $(T=512)$}
\begin{center}
\includegraphics[width=8cm, height=5cm]{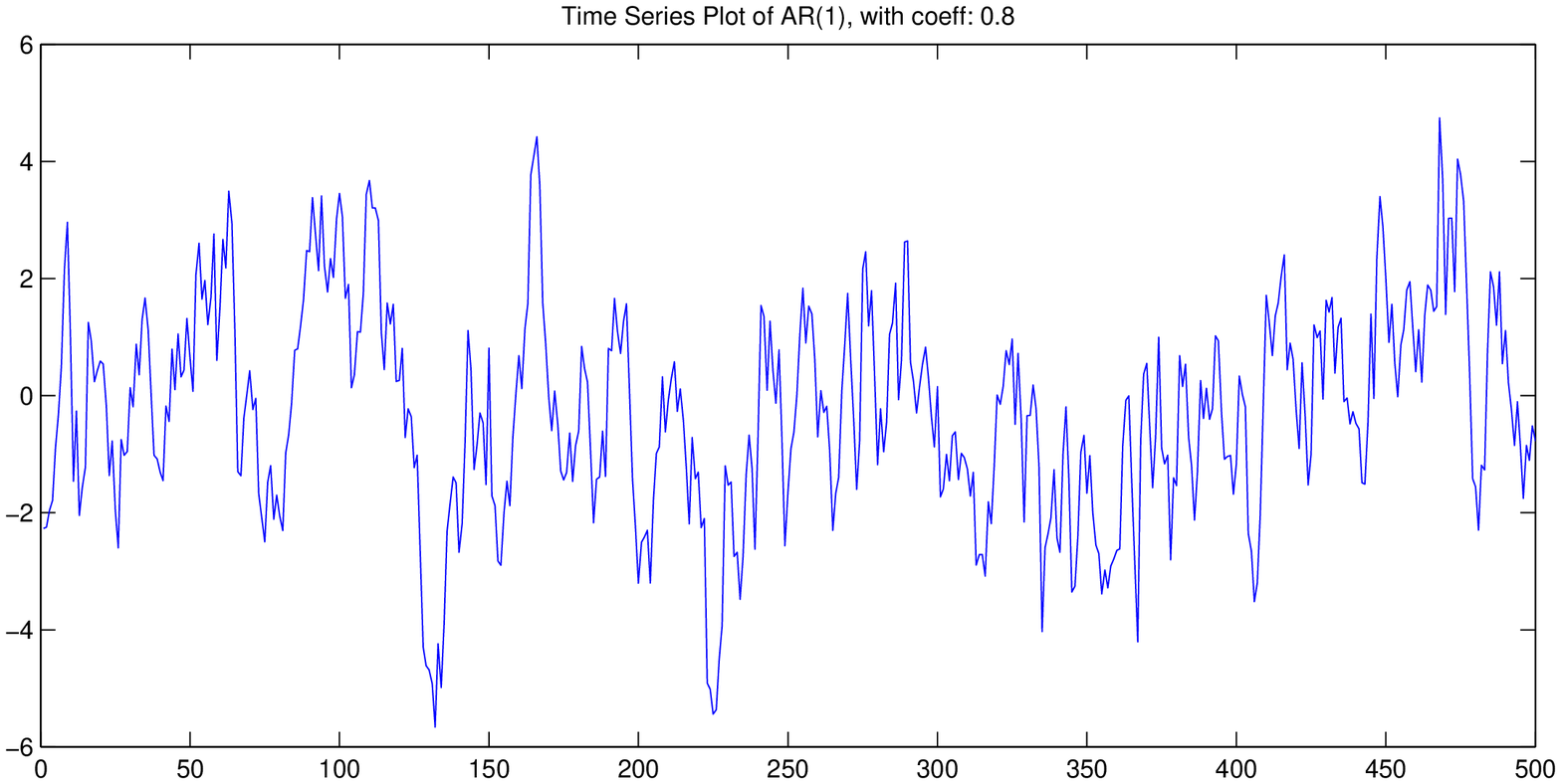}
\includegraphics[width=8cm, height=5cm]{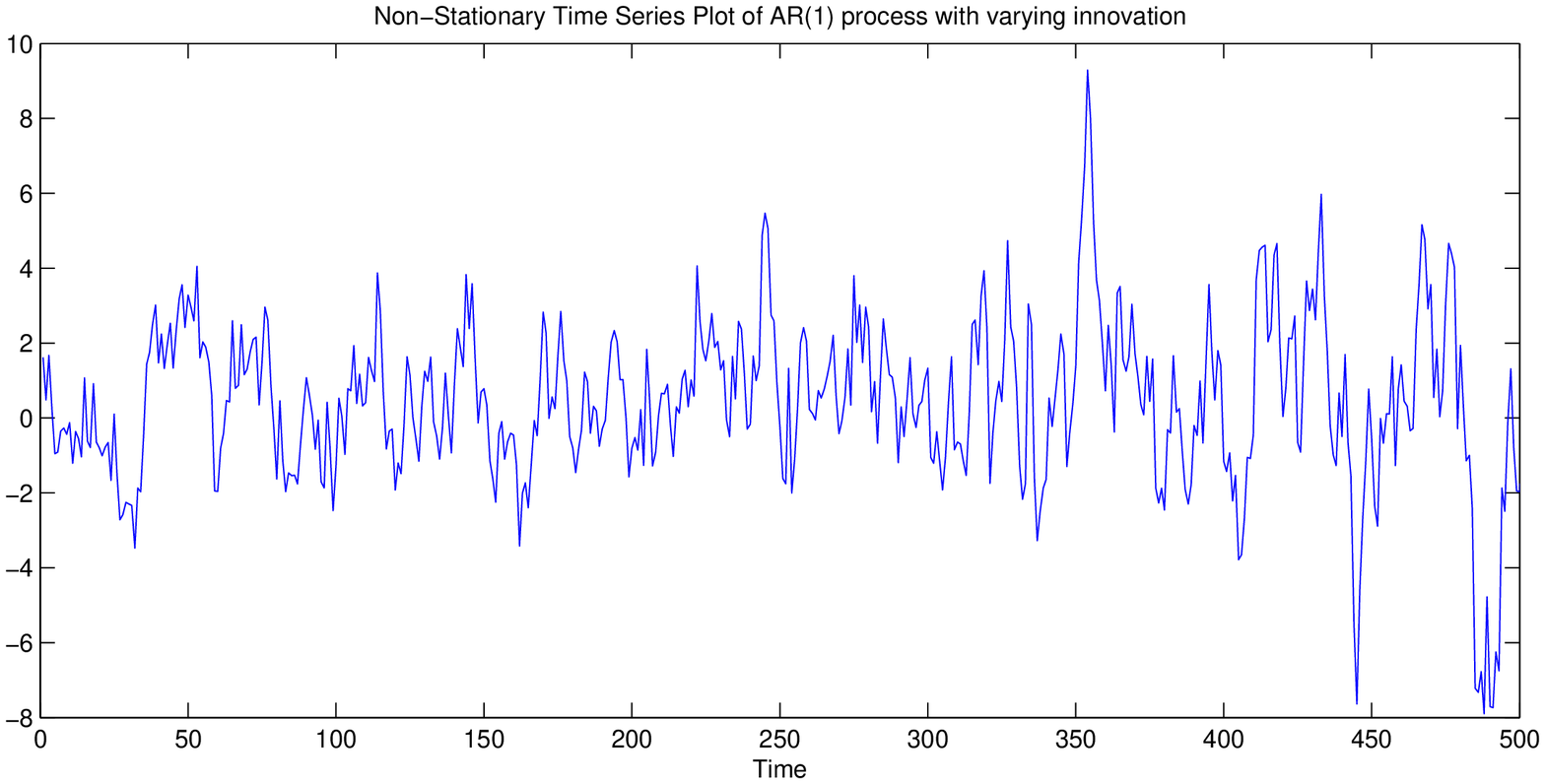}
\end{center}
\caption{\label{fig:3} Left: Stationary realisation Model 1. Right: 
Nonstationary realisation Model 4.}
\end{figure}

We first investigate the behaviour of the test statistic under the 
null hypothesis of stationarity. 
\begin{itemize}
\item[(i)] Model 1: $X_{t} = 0.8X_{t-1} + \varepsilon_{t}$, where
$\{\varepsilon_{t}\}$ are iid Gaussian random variables. 
We do the test for consecutive lags $r=1,\ldots,m$, the results can be 
found in the table below.
A plot of the estimated finite sample density of the
test statistic is given in Figure \ref{fig:1}. 
\begin{table}[h!]
\hspace{20mm}\begin{tabular}{|l|l|l|l|}
\hline
$T=256$ & $m=1$ & $m=5$ & $m=10$ \\ 
\hline
\% reject & 6 & 5 & 8 \\ 
\hline
\end{tabular}
\begin{tabular}{|l|l|l|l|}
\hline
$T=512$ & $m=1$ & $m=5$ & $m=104$ \\ 
\hline
\% reject      & 5.3 & 6.4 & 6.5 \\
\hline
\end{tabular}
\end{table}
\item[(ii)]  Model 2:
$X_{t}-X_{t-1}-0.7X_{t-2}=\varepsilon _{t}+0.3\varepsilon_{t-1}+
2\varepsilon _{t-3}$, $\{\varepsilon _{t}\}$ are Gaussian. 
We do the test for consecutive lags $r=1,\ldots,m$, the results can be found in the table below. 
A plot of the estimated finite sample density of the 
test statistic $\mathcal{T}_{10}$ is given in Figure \ref{fig:1}.
\begin{table}[h!]
\hspace{20mm}\begin{tabular}{|l|l|l|l|}
\hline
$T=256$ & $m=1$ & $m=5$ & $m=10$ \\ 
\hline
\% reject & 4.9 & 5.9 & 6.3 \\ 
\hline
\end{tabular}
\begin{tabular}{|l|l|l|l|}
\hline
$T=512$ & $m=1$ & $m=5$  & $m=10$ \\ 
\hline
\% reject & 8.33 & 4.67 & 3.33 \\ 
\hline
\end{tabular}
\end{table}
\end{itemize}
We observe that under the null of stationarity,  
the percentage rejects, in the tables, and the plots of the
empirical density in Figure \ref{fig:1} suggest that the $\chi^{2}$-distribution 
approximates well the distribution of the test statistic $\mathcal{T}_{m}$.  


\subsection{Nonstationary time series}

\begin{figure}[h!]
\begin{center}
\includegraphics[width=12cm, height=5cm]{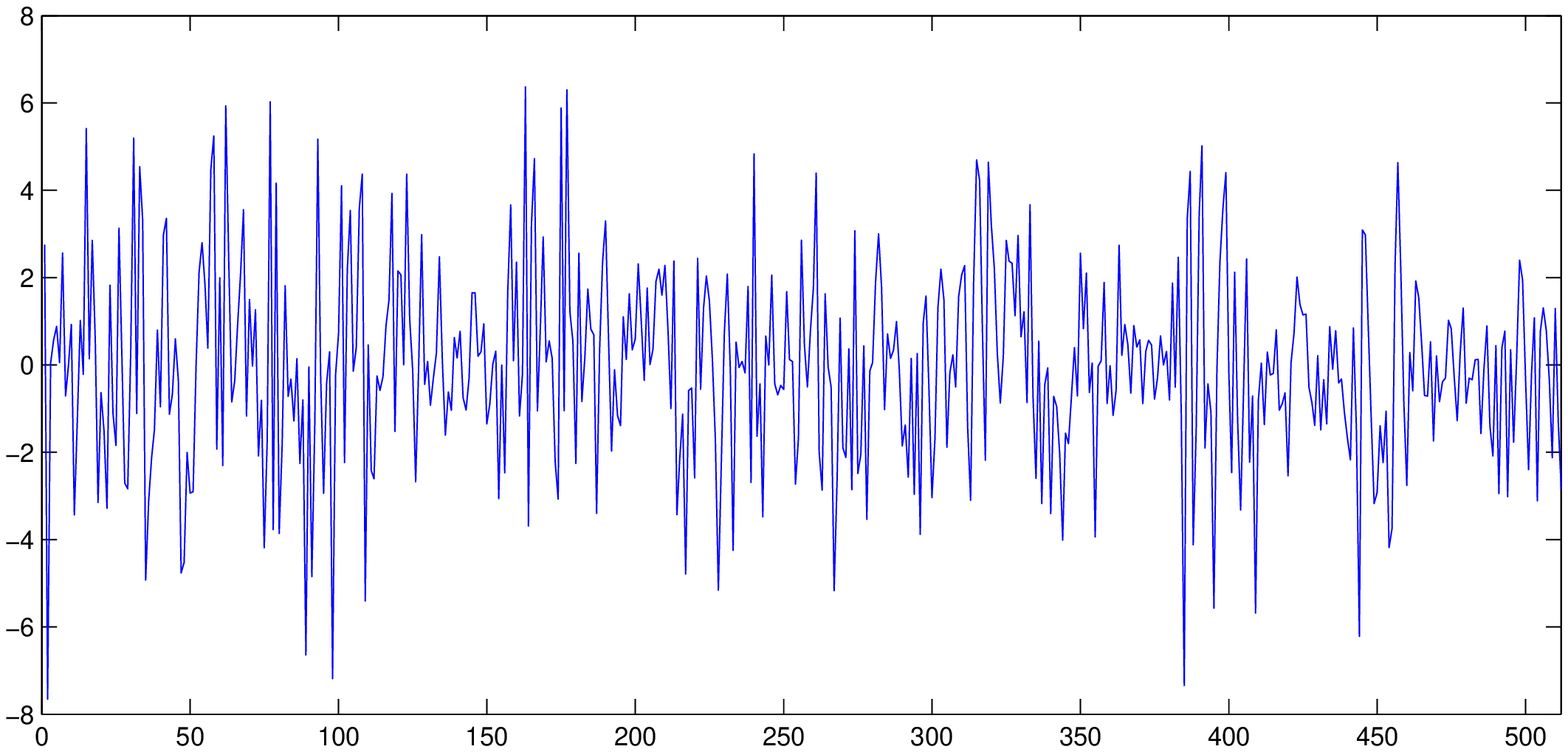}
\includegraphics[width=12cm, height=5cm]{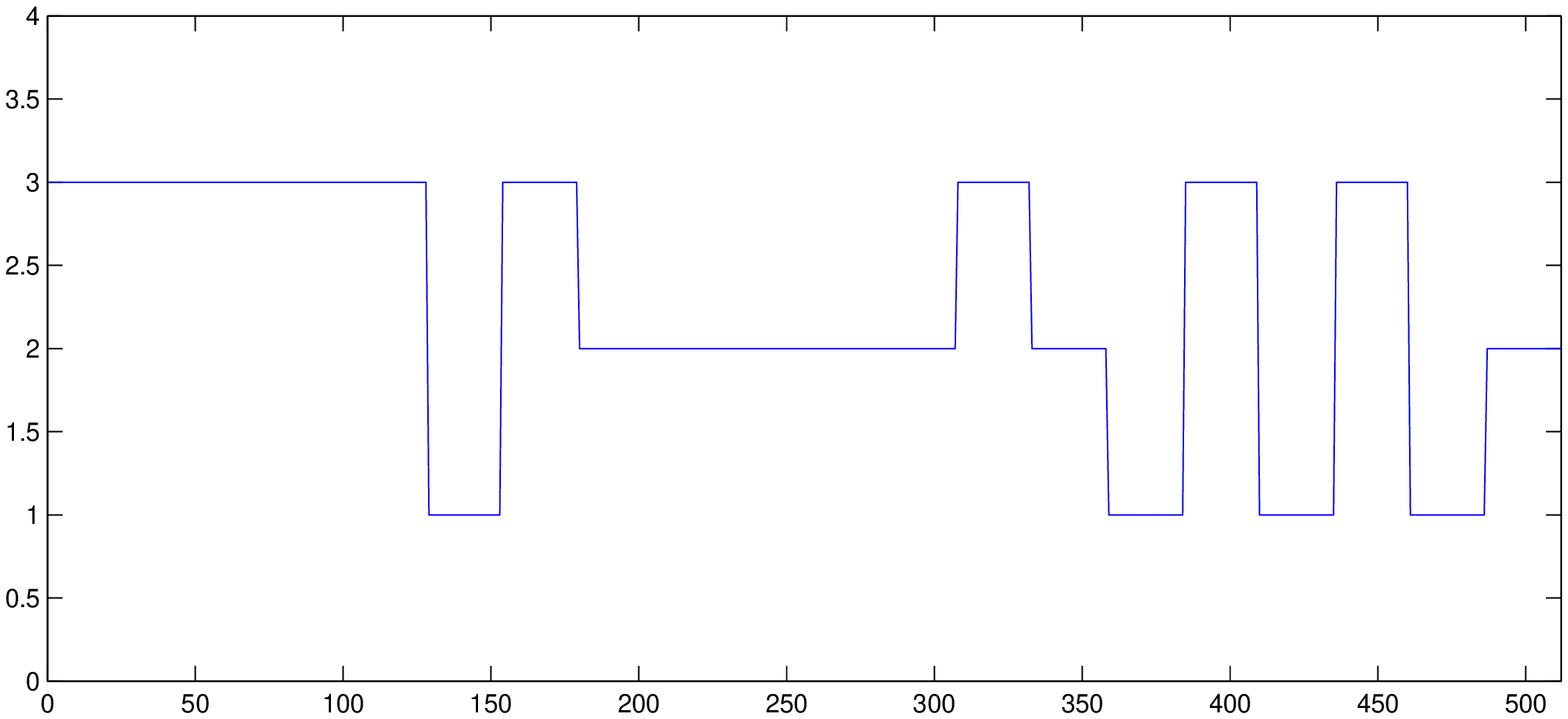}
\end{center}
\caption{\label{fig:4} Top: A realisation from nonstationary Model 6. Lower:
The time varying function $\sigma(\cdot)$}
\begin{center}
\includegraphics[width=12cm, height=8cm]{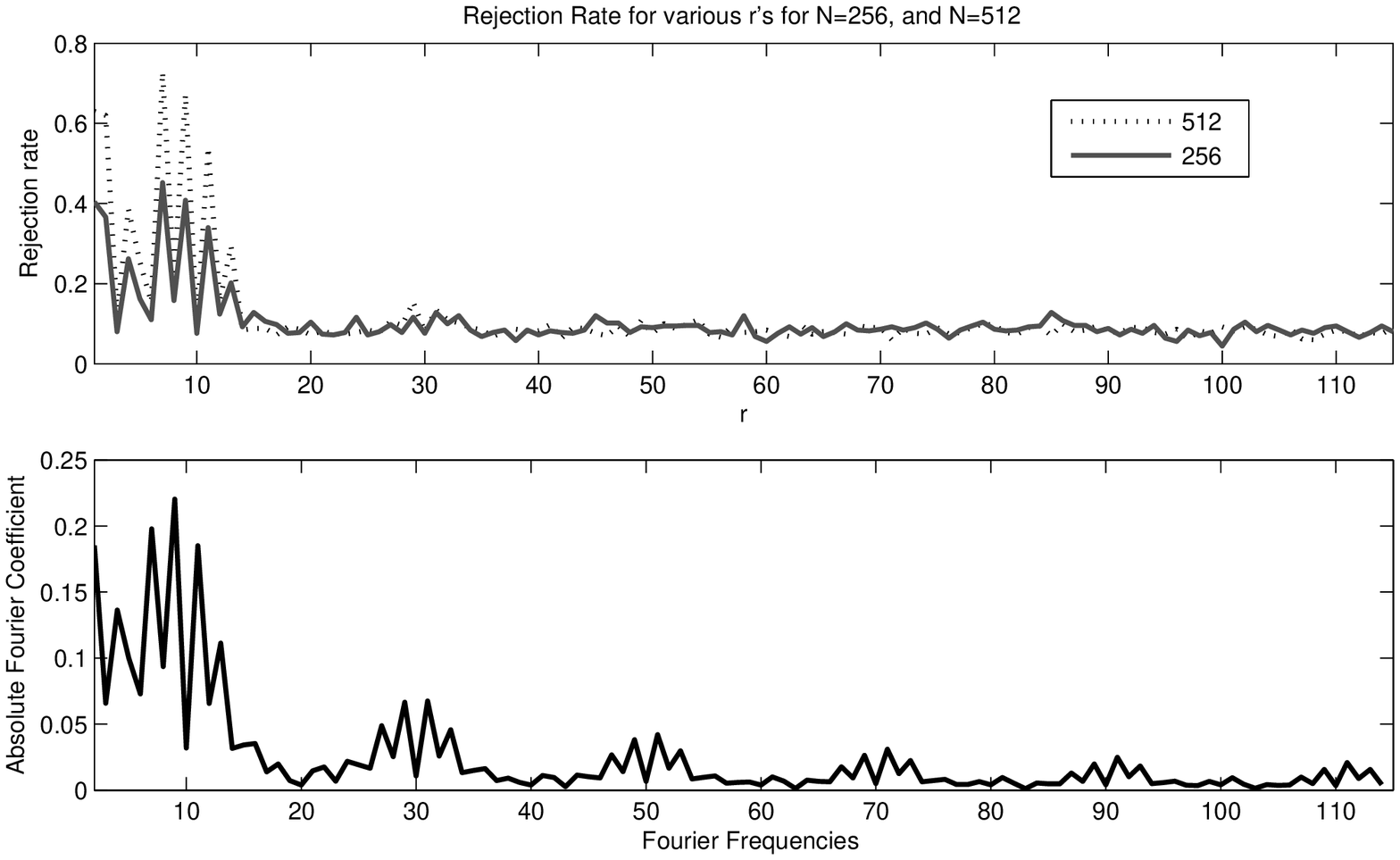}
\end{center}
\caption{\label{fig:5} Top: The x-axis is the lag used in the test statistic $r=1,\ldots,120$ and 
the $y$-axis is the corresponding rejection rate using the test statistic 
$\mathcal{T}_{1}> \chi^{2}_{2}(0.05)$ for 
$T=256$ and $T=512$. Lower: Absolute values of the Fourier coefficients of $\sigma(\cdot)$.}
\end{figure}

We now investigate the performance of the test statistic under different
types of nonstationary behaviour. 
\begin{itemize}
\item[(i)] Model 3: 
 In this model there is a change point occuring in the later part
of the time series: 
$X_{t}=1.5X_{t-1}-0.75X_{t-1}+\varepsilon _{t}$ for  
$t=1,\ldots,0.75 T$ and 
$X_{t}=0.8X_{t-1}+\varepsilon _{t}$  for $t=(0.75T+1),\ldots,T$. 
We do the test for consecutive lags $r=1,\ldots,m$, the results can be found in the table below.
A plot of the estimated finite sample density of the 
test statistic $\mathcal{T}_{10}$ is given in Figure \ref{fig:2}.
{\small \begin{table}[h!]
\hspace{20mm}\begin{tabular}{|l|l|l|l|}
\hline
$T=256$ & m=1 & m=5 & m=10 \\ 
\hline
\% reject & 100 & 100 & 100 \\ 
\hline
\end{tabular}
\begin{tabular}{|l|l|l|l|}
\hline
$T=512$ & m=1 & m=5 & m=10 \\ 
\hline
\% reject & 100 & 100 & 100 \\ 
\hline
\end{tabular}
\end{table}}

\item[(ii)] Model 4: In this model the variance of the innovation changes smoothly 
over time: 
$X_{t}=0.8X_{t-1}+\sigma _{t}\varepsilon _{t},$ where $\sigma _{t}=
(\frac{1}{2}+
\sin (\frac{2t\pi }{512})+0.3\cos (\frac{2t\pi }{512}))$, where 
$\{\varepsilon _{t}\}$ are Gaussian. 
We do the test for consecutive lags $r=1,\ldots,m$, the results can be found in the table below.
A plot of the estimated finite sample density of the
test statistic $\mathcal{T}_{10}$ is given in Figure \ref{fig:2}.
{\small \begin{table}[h!]
\hspace{20mm}\begin{tabular}{|l|l|l|l|}
\hline
T=256 & m=1 & m=5 & m=10 \\ 
\hline
\% reject & 6.6 & 16.9 & 26.6 \\ 
\hline
\end{tabular}
\begin{tabular}{|l|l|l|l|}
\hline
T=512 & m=1 & m=5 & m=10 \\ 
\hline
\% reject & 99.9 & 98 & 94.6 \\ 
\hline
\end{tabular}
\end{table}}

We mention, that for $1\leq t\leq 256$, function $\sigma_{t}$ defined above, does not vary 
much, which explains why the rejection rate for $T=256$ is relatively low. 
\item[(iii)] Model 5: In this model there is a change point, but the change is quite small: 
$X_{t}= 0.8X_{t-1}+\varepsilon _{t}$ for  
$t=1,\ldots,0.5 T$ and 
$X_{t}=0.6X_{t-1}+\varepsilon _{t}$  for $t=(0.5T+1),\ldots,T$, 
  where $\{\varepsilon _{t}\}$ are Gaussian.
We do the test for consecutive lags $r=1,\ldots,m$, the results can be found in the table below.
\begin{table}[h!]
\hspace{20mm}\begin{tabular}{|l|l|l|l|}
\hline
T=256 & m=1 & m=5 & m=10 \\ 
\hline
\% reject & 52 & 35.6 & 25.2 \\ 
\hline
\end{tabular}
\begin{tabular}{|l|l|l|l|}
\hline
T=512 & m=1 & m=5 & m=10 \\ 
\hline
\% reject & 82 & 66 & 52 \\ 
\hline
\end{tabular}
\end{table}
\item[(iv)]  Model 6: In this model, the time series is independent with time-varying variance.
Define the piecewise varying function $\sigma:[0,1]\rightarrow \mathbb{R}$
\begin{eqnarray*}
\sigma(u) = 
\left\{
\begin{array}{cl}
 1 & \textrm{ for } u\in\{[\frac{5}{20},\frac{6}{20}),[\frac{14}{20},\frac{15}{20}),
[\frac{16}{20},\frac{17}{20}),[\frac{18}{20},\frac{19}{20})\}\\
 2 &\textrm{ for } u \in\{[\frac{8}{20},\frac{12}{20}),[\frac{13}{20},\frac{14}{20}), 
[\frac{19}{20},1]\} \\
 3 & \textrm{ for } u\in\{[0,\frac{5}{20}), [\frac{6}{20},\frac{7}{20}),
[\frac{12}{20},\frac{13}{20}),[\frac{15}{20},\frac{16}{20}),
[\frac{17}{20},\frac{18}{20})\}\\
\end{array}
\right.
\end{eqnarray*}
and the time series $X_{t,T} = \sigma(\frac{t}{T})\varepsilon_{t}$,
where $\{\varepsilon_{t}\}$ are iid Gaussian random variables. 
A plot of a realisation of the time series and the function $\sigma(\cdot)$
is given in Figure \ref{fig:4}. 
We do the test for consecutive lags $r=1,\ldots,m$, the results can be found in the table below.
\begin{table}[h!]
\hspace{20mm}\begin{tabular}{|l|l|l|l|}
\hline
T=256 & m=1 & m=5 & m=10 \\ 
\hline
\% reject & 39 & 59 &  79 \\ 
\hline
\end{tabular}
\begin{tabular}{|l|l|l|l|}
\hline
T=512 & m=1 & m=5 & m=10 \\ 
\hline
\% reject & 62  & 85  &  99 \\ 
\hline
\end{tabular}
\end{table}

We observe that the rejection rate increases as 
the number of lags used in the test grows. We now investigate why. 
In Figure \ref{fig:5}, we plot the rejection rate of $\mathcal{T}_{1}$
(hence one lag) at lags $r=1,\ldots,120$, we do 
this for both sample sizes $T=256$ and $T=512$ and we 
also plot the absolute values of the Fourier coefficients of the function 
$\sigma(\cdot)$ for $r=1,\ldots,120$ (ie. $\{|a_{r}|\}$, where 
$\sigma(u) = \sum_{r}a_{r}\exp(-ir\omega)$, which are estimated with 
$\hat{a}_{r,T} = \frac{1}{T}\sum_{t=1}^{512}\sigma(\frac{t}{512})\exp(\frac{-i2\pi r}{512})$). 
We observe that with one lag (in the test statistic) 
the rejection rate is greatest at the frequencies $r$ that the 
Fourier coefficients are largest. This illustrates
well the theory of the test statistic under the alternative. 
More precisely, 
for this model, the time-varying spectral density is approximately 
$f(u,\omega) = \sigma(u)^{2}$, and the noncentrality parameter
is largest when the Fourier coefficient
$\int_{0}^{1}f(u,\omega)\exp(-2i\pi r u)du = \int_{0}^{1}\sigma(u)^{2}\exp(-2i\pi r u)du$
is largest. 
\end{itemize}
We observe that for various types of nonstationary behaviour the test has good power. 
Moreover, we observe that for many types of nonstationarity,  
by using a few small number of lags $r$, we are still  able to reject the null hypothesis.

\section{Real data analysis}\label{sec:data}

To illustrate the test for second order 
stationarity we consider two real data examples. 
For both data sets we will use the test statistic
$\mathcal{T}_{4} = T\sum_{r=1}^{4}|\hat{c}_{T}(r)|^{2}$. 
We choice $m=4$, because the simulations in the previous
section show that most nonstationary behaviour appears to be 
captured in the first four lags. We recall
that under the null $\mathcal{T}_{4}$ asymptotically has a chi squared 
distribution with eight degrees of freedom. 

\begin{figure}[h!]
\begin{center}
\includegraphics[width=12cm, height=5cm]{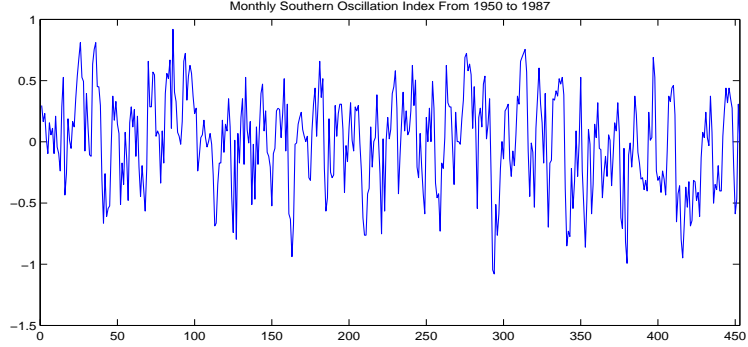}
\end{center}
\caption{Monthly southern oscillation
index time series between January 1950 to 
December 1987 \label{fig:6}} 
\end{figure}

We first test for second order stationarity of the monthly southern oscillation
index time series observed between January 1950 to 
December 1987 ($T=453$). The data can be found at 
{\tt http://www.stat.pitt.edu/stoffer/tsa2/}, a plot is given in 
Figure \ref{fig:6}. 
The test statistic gives the value
$\mathcal{T}_{4} = 2.66$, which corresponds to a p-value of
$0.95$, hence there is no evidence to reject the null of second 
order stationarity.

\begin{figure}[h!]
\begin{center}
\includegraphics[width=12cm, height=5cm]{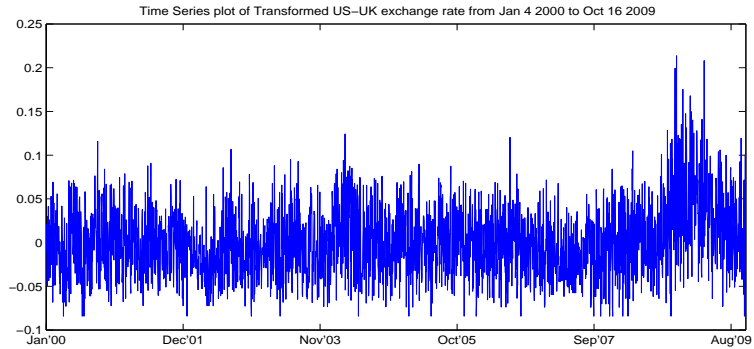}
\end{center}
\caption{Daily British pound/US dollar
exchange rate data observed between January 2000 to 
October 2009\label{fig:7}} 
\end{figure}

In our second data example we consider the daily British pound/US dollar
exchange rate data observed between January 2000 to 
October 2009. This data was obtained from
{\tt http://federalreserve.gov/releases/h10/Hist}. In order to ensure
the existence of moments we tranformed the data and 
considered the square root of the absolute log differences, that is 
$X_{t} = |\log Y_{t}^{2} - \log Y_{t-2}^{2}|^{1/2}$, where 
$Y_{t}$ is the exchange rate at time $t$. 
A plot of the transformed data is given in Figure \ref{fig:7}. 
The test statistic based on the 
entire data set gave the value 
$\mathcal{T}_{4}=99.4$, which corresponds to the   
p-value $p\approx 0$. Thus suggests there is evidence to reject the 
null of second order stationarity. To locate the regions of nonstationarity
the data is partioned into segments of half length, quarter length and 
eighth length and the test was performed on each of these segments. The 
results are given in Table \ref{table:Ex}.  Studying 
Table \ref{table:Ex} and the p-values, there is evidence to 
suggest that for most periods there is no evidence to reject the null
of stationarity. However, over the blocks June 2002 - November 2004,  
and August 2008 - October 2009, the data seems to be second order nonstationary.
This information can be used to fit a model with time-varying parameters to 
the data.

\begin{table}[h]
\begin{center}

\begin{tabular}{|p{1.75cm}|p{1.75cm}|p{1.75cm}|p{1.75cm}|p{1.75cm}|p{1.75cm}|p{1.75cm}|p{1.75cm}|}
\begin{sideways}\parbox{10mm}{Jan'00} \end{sideways}
& \begin{sideways}\parbox{10mm}{Mar'01} \end{sideways}
& \begin{sideways}\parbox{10mm}{Jun'02} \end{sideways}
& \begin{sideways}\parbox{10mm}{Sep'03} \end{sideways}
& \begin{sideways}\parbox{10mm}{Nov'04} \end{sideways}
& \begin{sideways}\parbox{10mm}{Feb'06} \end{sideways}
& \begin{sideways}\parbox{10mm}{May'07} \end{sideways}
& \begin{sideways}\parbox{10mm}{Aug'08} \end{sideways}\\\hline
\multicolumn{8}{|c|}{$T=99.4$  $p=0.000$}\\\hline
\multicolumn{4}{|c|}{$T=29.5$  $p=0.000$}
&\multicolumn{4}{|c|}{$T=78.43$ $p=0.000$}\\
\hline
\multicolumn{2}{|p{3.5cm}|}{$T=20.0$ $p=0.010$}
&\multicolumn{2}{|p{3.5cm}|}{$T=23.9$ $p=0.002$}
&\multicolumn{2}{|p{3.5cm}|}{$T=5.1$  $p=0.752$}
&\multicolumn{2}{|p{3.5cm}|}{$T=51.4$ $p=0.000$}  \\
\hline
$T=7.2$ $p=0.520$ & $T=15.5$ $p=0.055$ & $T=19.4$ $p=0.013$ & $T=18.2$ 
$p=0.020$ & $T=5.5$ $p=0.702$ & $T=12.0$ $p=0.151$ 
& $T=12.1$ $p=0.145$ & $T=17.9$ $p=0.022$
\\\hline
\end{tabular}
\caption{\label{table:Ex}
Test Statistics and p-value}
\end{center}
\end{table}


\section{Conclusions}

In this article we have considered a test for second order stationarity. 
The test is based on the property that the DFT of a second order 
stationary time series is 
close to uncorrelated. The sampling properties of the test statistic 
under the null are derived under the assumption that the time series 
statisfies a short memory, $MA(\infty)$ representation. However, 
empirical evidence suggest that 
similar sampling properties also hold in the case of both stationary 
long memory and nonlinear time series too. 
For this general case, it may be possible to use some of the theory 
developed in \citeA{p:kok-mik-00}, however, this is beyond the scope of the current paper
and is future work.

\subsection*{Acknowledgements}

The authors are grateful to Professor Manfred Deistler for several interesting
discussions. 
This work has been partially supported by the NSF grant DMS-0806096.

\newpage

\newpage

\appendix

\section{Appendix}

In this appendix we prove the results from the main section. 

For short hand, when it is clear that $T$ plays a role, we use 
the notation
$J_{k}: = J_{T}(\omega_{k})$, $\bar{J}_{k} =  J_{T}(-\omega_{k})$, 
$\hat{f}_{k}: = \hat{f}_{T}(\omega_{k})$ and $f_{k} = f(\omega_{k})$. 

\subsection{Some results on DFTs and Fourier coefficients}\label{sec:DFTA}

In the sections below, under various assumptions of the dependence
of $\{X_{t}\}$, we will show asymptotic normality and obtain the 
mean and variance of $\widetilde{c}_{T}(r)$. 
In the case that 
$\{X_{t}\}$ is a short memory, stationary time series, then 
it is relatively straightforward to evaluate the variance of 
$\widetilde{c}_{T}(r)$, since
$\{J_{T}(\omega_{k})\overline{J_{T}(\omega_{k+r})}\}$ are close to 
uncorrelated random variables.
However, in the nonstationary case this no longer holds and we have to use 
some results in Fourier analysis to derive 
$\var(\widetilde{c}_{T}(r))$. To do this 
we start by studying the general random variable
\begin{eqnarray*}
H_{T} = \frac{1}{T}\sum_{k=1}^{T}H(\omega_{k})J_{T}(\omega_{k})
\overline{J_{T}(\omega_{k+r})}, 
\end{eqnarray*}
where $J_{T}(\omega_{k}) = 
\frac{1}{\sqrt{2\pi T}}\sum_{t=1}^{T}X_{t}\exp(\frac{2\pi i tk}{T})$.
We will show that under certain conditions on $H(\cdot)$, $H_{T}$
can be written as the weighted average of $X_{t}$. 
Expanding $J_{T}(\omega_{k})\overline{J_{T}(\omega_{k})}$ we see that 
$H_{T}$ can be written as 
\begin{eqnarray*}
H_{T} = \frac{1}{T}\sum_{t,\tau}X_{t}X_{\tau}\exp(-i\omega_{r}\tau) 
\bigg(\frac{1}{T}
\sum_{k=1}^{T}H(\omega_{k})\exp(i\omega_{k}(t-\tau))\bigg). 
\end{eqnarray*}
Without any smoothness assumptions on $H(\omega_{k})$, it is not clear 
whether the inner sum of the above converges to zero as $|t-\tau|\rightarrow \infty$,
and if the variance of $H_{T}$ converges to zero. 
However, let us suppose that $\sup_{\omega}|H^{\prime}(\omega)|<\infty$.  
In this case, the DFT of $H(\omega)$, 
$\frac{1}{T}\sum_{k=1}^{T}H(\omega_{k})\exp(i\omega_{k}(t-\tau))$,  
is an approximation of 
the Fourier coefficient $
h(t-\tau) = \int H(\omega)\exp(i(t-\tau)\omega)d\omega$
(the error in this approximation is discussed below). Noting that 
the Fourier coefficients $h(k)\rightarrow 0$ as $k\rightarrow \infty$, 
we have 
\begin{eqnarray*}
H_{T} \approx \frac{1}{T}\sum_{t=1}^{T}X_{t}
\sum_{\tau}h(t-\tau)X_{\tau}\exp(-i\omega_{r}\tau) := 
\frac{1}{T}\sum_{t=1}^{T}X_{t}Y_{t}^{(r)}. 
\end{eqnarray*}
Now under relatively weak conditions on the
Fourier coefficients, $\{h(k)\}$, $Y_{t}^{(r)}$ is almost surely finite (and second order stationary if 
$\{X_{t}\}$ were stationary). 
Hence $H_{T}$ can be considered a weighted average of $\{X_{t}\}$, where the
weights in this case
are random but almost surely finite. We now justify some of the approximations 
discussed above and state some well know results in Fourier analysis. 
An interesting overview of results in Fourier analysis
is given in \citeA{b:bri-hen-97}. 

The following theorem is well known, see for example, \citeA{b:bri-hen-97}, 
Theorem 6.2, for the proof. 
\begin{theorem}\label{thm:DFT-fourier}
Suppose that $g[0,\Omega]\rightarrow \mathbb{R}$ is a periodic function with 
period $\Omega$. We shall assume that either (a) 
$\sup_{x}|g^{\prime\prime}(x)| <\infty$
or (b) $\sup_{x}|g^{\prime}(x)| < \infty$ and 
$g^{\prime}(\cdot)$ is piecewise montone function.
Let 
\begin{eqnarray*}
a(s) = \frac{1}{\Omega}\int_{0}^{\Omega} 
g(x)\exp(isx)dx \quad \textrm{ and } \quad  
a_{T}(s) = \frac{1}{T}\sum_{k=1}^{T} g(\frac{\Omega k}{T})
\exp(i\frac{\Omega k}{T}).
\end{eqnarray*} 
Therefore for all $s$, we have in case (a)
$|a(s)|\leq C\sup_{x}|g^{\prime\prime}(x)|s^{-2}$ 
and in case (b) $|a(s)|\leq C\sup_{x}|g^{\prime}(x)|s^{-2}$, 
where $C$ is constant independent of $s$ and $g(\cdot)$. 

Moreover $\sup_{1\leq s\leq T}|a(s)-a_{T}(s)| \leq CT^{-2}$. 
\end{theorem}

We now apply the above result to our setup. 
We will use Lemma \ref{lemma:GDFT}, below, to prove the asymptotic
normality result in Section \ref{sec:asym-normality}. 
\begin{lemma}\label{lemma:GDFT}
Suppose Assumption \ref{assum:stat} or  Assumption \ref{assum:nonstat} is 
satisfied. And let 
$f(\omega)$ be the spectral density of the stationary linear time series or 
the integrated spectral density of the locally stationary time series. Let
\begin{eqnarray}
\label{eq:fourierCOA}
G_{T,\omega_{r}}(s) &=& \frac{1}{T}
\sum_{k=1}^{T}\frac{1}{(f(\omega_{k})
f(\omega_{k} + \omega_{r}))^{1/2}}\exp(is\omega_{k}) \\
\textrm{ and }
G_{\omega_{r}}(s) &=& \frac{1}{2\pi}\int_{0}^{2\pi} \frac{1}{(f(\omega) 
f(\omega + \omega_{r}))^{1/2}} 
\exp(is\omega)d\omega. \label{eq:fourierCO}
\end{eqnarray}
Under the stated assumptions we have either 
$(f(\omega)f(\omega + \omega_{r}))^{-1/2}$ has a bounded second derivative
(under Assumption \ref{assum:stat}(iv-a) or Assumption \ref{assum:nonstat})
or that $(f(\omega)f(\omega + \omega_{r}))^{-1/2}$ has a bounded first derivative and 
is piecewise monotone (under Assumption \ref{assum:stat}(iv-b)). 
Therefore, $\sup_{\omega_{r}}\sum_{s} |G_{\omega_{r}}(s)| <\infty$ and 
$\sup_{\omega_{r}}\sum_{s=1}^{T} |G_{T,\omega_{r}}(s)| <\infty$. 
\end{lemma}
PROOF. The above is a straightforward application of Theorem 
\ref{thm:DFT-fourier}. \hfill $\Box$

\vspace{3mm}

We will use the result below in Sections \ref{sec:proof-diff} and 
\ref{sec:var}. 
\begin{lemma}\label{lemma:fourierstuff}
Suppose Assumption \ref{assum:nonstat} is satisfied. Define the $n$th order
cumulant spectra as 
\begin{eqnarray*}\label{eq:spectra-general}
f_{n}(u,\omega_{1},\ldots,\omega_{n-1}) = \frac{1}{(2\pi)^{(n/2)-1}}
\bigg\{\prod_{j=1}^{n-1}A(u,\omega_{j})\bigg\}A(u,-\sum_{j=1}^{n-1}\omega_{j}),
\end{eqnarray*}
and the Fourier transform 
\begin{eqnarray}
\label{eq:Fnomega}
F_{n}(k;\omega_{1},\ldots,\omega_{n-1}) = \int_{0}^{1}f_{n}(u,\omega_{1},\ldots,\omega_{n-1})
\exp(i2\pi ku)du. 
\end{eqnarray}
\begin{itemize}
\item[(i)] If Assumption \ref{assum:nonstat}(iv)(a) holds, then 
$\sup_{u,\omega_{1},\ldots,\omega_{n}}|\frac{\partial^{2}f_{n}(u,\omega_{1},\ldots,\omega_{n-1})}{\partial u^{2}}|<
\infty$
and 
\begin{eqnarray}\label{eq:Fbounda}
\sup_{\omega_{1},\ldots,
\omega_{n-1}}|F_{n}(k;\omega_{1},\ldots,\omega_{n-1})| \leq 
C\sup_{u,\omega_{1},\ldots,\omega_{n}}|\frac{\partial^{2}
f_{n}(u,\omega_{1},\ldots,\omega_{n-1})}{\partial u^{2}}|\frac{1}{|k|^{2}}.
\end{eqnarray}
\item[(ii)] If Assumption \ref{assum:nonstat}(iv)(b) holds, then 
$\sup_{u,\omega_{1},\ldots,\omega_{n}}|\frac{\partial f_{n}(u,\omega_{1},\ldots,\omega_{n-1})}{\partial u}|<\infty$,
$\frac{\partial f_{n}(u,\omega_{1},\ldots,\omega_{n-1})}{\partial u}$ is a piecewise monotone function in 
$u$ and 
\begin{eqnarray}
\label{eq:Fboundb}
\sup_{\omega_{1},\ldots,\omega_{n}}|F_{n}(k;\omega_{1},\ldots,\omega_{n-1})| \leq 
C\sup_{u,\omega_{1},\ldots,\omega_{n}}|\frac{\partial f_{n}(u,\omega_{1},\ldots,\omega_{n-1})}{
\partial u}|\frac{1}{|k|^{2}}.
\end{eqnarray}
\end{itemize}
We note that the
 constant $C$ is independent of the function $f_{n}(\cdot)$ and $k$.
\end{lemma}
PROOF. To ease notation we only prove (ii), the proof of (i) is similar. 
We note that 
\begin{eqnarray*}
(2\pi)^{(n/2)-1}\frac{d f_{n}(u,\omega_{1},\ldots,\omega_{n-1})}{du} &=& 
A(u,-\sum_{j=1}^{n-1}\omega_{j})\sum_{r=1}^{n-1}
\frac{d A(u,\omega_{r})}{du}\prod_{j\neq r} A(u,\omega_{j}) + \\
 && \frac{\partial A(u,-\sum_{j=1}^{n-1}\omega_{j})}{\partial u}
\prod_{j=1}^{n-1} A(u,\omega_{j}). 
\end{eqnarray*}
Now under Assumption \ref{assum:nonstat}, $\sup_{u,\omega}|A(u,\omega)|$
and $\sup_{u,\omega}|\frac{\partial A(u,\omega)}{\partial u}|$ are bounded 
function, hence we see from the above that 
$\sup_{u,\omega}|\frac{\partial  f_{n}(u,\omega_{1},\ldots,\omega_{n-1})}{\partial u}|$ 
is bounded. Moreover, by using Theorem \ref{thm:DFT-fourier} we have 
(\ref{eq:Fbounda}). The proof of (ii) is similar, and we 
omit the details. \hfill $\Box$

\vspace{3mm}
We observe that in the stationary case $A(u,\omega)\equiv A(\omega)$, then 
$F_{n}(k;\omega_{1},\ldots,\omega_{n-1}) = 0$ for $k\neq 0$.

\vspace{3mm}
In the following lemma we consider the error in approximation of the 
DFT with the Fourier coefficient. 

\begin{lemma}\label{lemma:FDFT}
Suppose Assumption \ref{assum:nonstat} is satisfied. Let 
$F_{n}(k;\omega_{1},\ldots,\omega_{n-1})$
be defined as in (\ref{eq:Fnomega}) and let 
\begin{eqnarray}
\label{eq:FTomega}
F_{n,T}(s;\omega_{1},\ldots,\omega_{n}) 
= \frac{1}{T}\sum_{t=1}^{T}f_{n}(\frac{t}{T},\omega_{1},
\ldots,\omega_{n-1})\exp(i2s\pi t/T).  
\end{eqnarray}
Then under  Assumption \ref{assum:nonstat}(v)(a) we have 
\begin{eqnarray*}
\sup_{\omega_{1},\ldots,\omega_{n-1}}
\big| F_{n,T}(s;\omega_{1},\ldots,\omega_{n})  - 
F_{n}(s;\omega_{1},\ldots,\omega_{n-1})
\big| \leq C\sup_{u,\omega_{1},\ldots,\omega_{n}}|\frac{\partial^{2}
f_{n}(u,\omega_{1},\ldots,\omega_{n-1})}{\partial u^{2}}|\frac{1}{T^{2}},
\end{eqnarray*}
and  under  Assumption \ref{assum:nonstat}(v)(b) we have 
\begin{eqnarray*}
\sup_{\omega_{1},\ldots,\omega_{n-1}}
\big| F_{n,T}(s;\omega_{1},\ldots,\omega_{n})  - F_{n}(s;\omega_{1},
\ldots,\omega_{n-1})
\big| \leq C\sup_{u,\omega_{1},\ldots,\omega_{n}}|
\frac{\partial^{}f_{n}(u,\omega_{1},\ldots,\omega_{n-1})}{\partial u^{}}|
\frac{1}{T^{2}},
\end{eqnarray*}
where $C$ is a constant independent of $f_{n}(\cdot)$. 
\end{lemma}
PROOF. The proof follows immediately from Theorem \ref{thm:DFT-fourier} and 
Lemma \ref{lemma:fourierstuff}. \hfill $\Box$

\subsection{Proof of Theorems \ref{lemma:cov-diff} and 
\ref{lemma:cov-diff-nonstat}}\label{sec:proof-diff}


The follow result is due to \citeA{p:pap-09}, Lemma 6.2, and is a 
generalisation of 
\citeA{b:bri-81}, Theorem 4.3.2, for locally stationary time series.  
\begin{lemma}[Paparoditis (2009), Lemma 6.2]\label{lemma:pap}
Suppose that Assumption \ref{assum:nonstat} holds and 
let $f_{n}(\cdot)$ be defined as in (\ref{eq:spectra-general}). Then for 
$\omega_{1},\ldots,\omega_{n}\in [0,2\pi]$ we have 
\begin{eqnarray*}
\cum(J_{T}(\omega_{1}),\ldots,J_{T}(\omega_{n})) = 
\frac{(2\pi)^{(n/2)-1}}{T^{n/2}}\sum_{t=1}^{T}
f_{n}(\frac{t}{T},\omega_{1},\ldots,\omega_{n-1})\exp(it\sum_{j=1}^{n}\omega_{j}) + 
O(\frac{(\log T)^{n-1}}{T^{n/2}}). 
\end{eqnarray*}
\end{lemma}

Now the following corollary immediately follows from 
Lemmas \ref{lemma:FDFT} and  \ref{lemma:pap}. 
\begin{corollary}\label{cor:boundcumulants}
Suppose that Assumption \ref{assum:nonstat} holds, and let 
$F_{n}(\cdot)$ be defined as in (\ref{eq:Fnomega}). Then we have 
\begin{eqnarray*}
\cum(J_{T}(\omega_{j_{1}}),\ldots,J_{T}(\omega_{j_{n}})) = 
\frac{(2\pi)^{(n/2)-1}}{T^{(n/2)-1}}F_{n}(j_{1}+\ldots+j_{n};\omega_{j_{1}},\ldots,\omega_{j_{n-1}}) + 
O(\frac{(\log T)^{n-1}}{T^{n/2}} + \frac{1}{T^{2}})
\end{eqnarray*}
and 
\begin{eqnarray*}
\big|\cum(J_{T}(\omega_{j_{1}}),\ldots,J_{T}(\omega_{j_{n}}))\big| \leq 
C\frac{1}{T^{(n/2)-1}|j_{1}+\ldots+j_{n}|} + \frac{C(\log T)^{n-1}}{T^{n/2}} + \frac{C}{T^{2}},
\end{eqnarray*}
where $\omega_{j_{r}} = 2\pi j_{r}/T$, $-T \leq  j_{r}\leq T$ 
and $C$ is constant independent of $\omega$.
\end{corollary}
The following well known result represents moments in terms of cumulants. 
\begin{lemma}\label{lemma:zubrenko}
Let us suppose that $\sup_{t}\Ex(|X_{t}|^{n})<\infty$. Then we have 
\begin{eqnarray}
\Ex(X_{t_{1}},\ldots,X_{t_{n}}) = \sum_{\pi}
\prod_{B\in \pi}\cum(X_{i};i\in B),
\end{eqnarray}
where $\pi$ is a partition of $\{t_{1},t_{2},\ldots,t_{n}\}$ and the sum $\sum_{\pi}$ is done
over all partitions of $\{t_{1},t_{2},\ldots,t_{n}\}$. 
\end{lemma}

We use the above lemmas below. 

\begin{lemma}\label{lemma:difference}
Suppose either Assumption \ref{assum:stat} or 
 Assumption \ref{assum:nonstat} holds. Let $f(\cdot)$ be the integrated 
spectral density
(or the true spectral density in the case of stationarity). Then we have 
\begin{itemize}
\item[(i)] \vspace{-4mm}
\begin{eqnarray*}
\Ex\bigg(J_{k_{1}}\bar{J}_{k_{1}+r}\bar{J}_{k_{2}}J_{k_{2}+r}\big\{\hat{f}_{k_{1}}\hat{f}_{k_{1}+r}
- \Ex(\hat{f}_{k_{1}}\hat{f}_{k_{1}+r})\big\}\big\{\hat{f}_{k_{2}}\hat{f}_{k_{2}+r}
- \Ex(\hat{f}_{k_{2}}\hat{f}_{k_{2}+r})\big\}\bigg) 
=  \frac{C}{bT}\bigg(
\frac{1}{|k_{1}-k_{2}|^{2}} + \frac{\log T}{T}\bigg), 
\end{eqnarray*}
\item[(ii)] 
\begin{eqnarray*}
\Ex\bigg\{\hat{f}_{k}\hat{f}_{k+r}
- \Ex(\hat{f}_{k}\hat{f}_{k+r})\bigg\}^{4} 
\leq \frac{C}{(bT)^{2}}
\end{eqnarray*}
\item[(iii)] Suppose Assumption \ref{assum:stat}(i) holds, then we have  
$\Ex\big\{\hat{f}_{k}\hat{f}_{k+r}\big\} - f_{k}f_{k+r} = O(b + \frac{1}{bT})$. Suppose
Assumption \ref{assum:stat}(iv-a) or Assumption \ref{assum:nonstat} holds, then we have 
$\Ex\big\{\hat{f}_{k}\hat{f}_{k+r}\big\} - f_{k}f_{k+r} = O(b^{2} + \frac{1}{bT})$. 
\item[(iv)] Suppose  Assumption \ref{assum:nonstat} holds, then we have 
\begin{eqnarray*}
\Ex(J_{s_{1}}J_{s_{2}}J_{s_{3}}J_{s_{4}}) &\leq& 
C\big(\frac{1}{|s_{1}+s_{2}|^{2}} + 
\delta_{T}\big)\big(\frac{1}{|s_{3}+s_{4}|^{2}} + \delta_{T}\big) 
+C\big(\frac{1}{|s_{1}+s_{3}|^{2}} + 
\delta_{T}\big)\big(\frac{1}{|s_{2}+s_{4}|^{2}} + \delta_{T}\big)\\
 && +C\big(\frac{1}{|s_{1}+s_{4}|^{2}} + 
\delta_{T}\big)\big(\frac{1}{|s_{2}+s_{3}|^{2}}) + \delta_{T}\big)\\
 && + \frac{C}{T}\frac{1}{|s_{1}+s_{2}+s_{3}+s_{4}|^{2}} 
 + O\big(\frac{(\log T)^{3}}{T^{2}} + \frac{1}{T^{2}} \big),
\end{eqnarray*}
where $\delta_{T} = O(\frac{\log T}{T} + \frac{1}{T^{2}})$. 
\end{itemize}
\end{lemma}
PROOF. To simplify notation in the proof we 
let $w_{j} = \frac{1}{bT}K(\frac{\omega_{j}}{b})$. 
By expanding the expectation in (i) we have 
\begin{eqnarray}
&& \Ex\bigg(J_{k_{1}}\bar{J}_{k_{1}+r}\bar{J}_{k_{2}}J_{k_{2}+r}\big\{\hat{f}_{k_{1}}\hat{f}_{k_{1}+r}
- \Ex(\hat{f}_{k_{1}}\hat{f}_{k_{1}+r})\big\}\big\{\hat{f}_{k_{2}}\hat{f}_{k_{2}+r}
- \Ex(\hat{f}_{k_{2}}\hat{f}_{k_{2}+r})\big\}\bigg) \nonumber\\
&=& \sum_{j_{1},j_{2},j_{3},j_{4}}w_{j_{1}}w_{j_{2}}w_{j_{3}}w_{j_{4}} \nonumber\\
&& \bigg\{\Ex(J_{k_{1}-j_{1}}\bar{J}_{k_{1}-j_{1}}J_{k_{1}+r-j_{2}}\bar{J}_{k_{1}+r-j_{2}}
J_{k_{1}}\bar{J}_{k_{1}+r}\bar{J}_{k_{2}}J_{k_{2}+r}J_{k_{2}-j_{3}}\bar{J}_{k_{2}-j_{3}}
J_{k_{2}+r-j_{4}}\bar{J}_{k_{2}+r-j_{4}}) \nonumber\\
&& - \Ex(J_{k_{1}-j_{1}}\bar{J}_{k_{1}-j_{1}}
J_{k_{1}+r-j_{2}}\bar{J}_{k_{1}+r-j_{2}})
\Ex(J_{k_{1}}\bar{J}_{k_{1}+r}\bar{J}_{k_{2}}J_{k_{2}+r}J_{k_{2}-j_{3}}\bar{J}_{k_{2}-j_{3}}
J_{k_{2}+r-j_{4}}\bar{J}_{k_{2}+r-j_{4}})\nonumber\\
&& - \Ex(J_{k_{2}-j_{3}}\bar{J}_{k_{2}-j_{3}}J_{k_{2}+r-j_{4}}\bar{J}_{k_{2}+r-j_{4}})
\Ex(J_{k_{1}}\bar{J}_{k_{1}+r}\bar{J}_{k_{2}}J_{k_{2}+r}J_{k_{1}-j_{1}}\bar{J}_{k_{1}-j_{1}}
J_{k_{1}+r-j_{2}}\bar{J}_{k_{1}+r-j_{2}})  \nonumber\\
&&+ \Ex(J_{k_{1}-j_{3}}\bar{J}_{k_{1}-j_{3}}
J_{k_{1}+r-j_{4}}\bar{J}_{k_{1}+r-j_{4}})\Ex(J_{k_{1}}\bar{J}_{k_{1}+r}\bar{J}_{k_{2}}J_{k_{2}+r})
\Ex(J_{k_{2}-j_{3}}\bar{J}_{k_{2}-j_{3}}J_{k_{2}+r-j_{4}}\bar{J}_{k_{2}+r-j_{4}})\bigg\}.   \label{eq:expand}
\end{eqnarray}
To prove the result we first represent the above moments in terms of 
cumulants using Lemma \ref{lemma:zubrenko}.
We observe that many of the terms will cancel, however those 
that do remain will involve at least one cumulant which has 
elements belong to the set 
$\{J_{k_{1}-j_{1}},\bar{J}_{k_{1}-j_{1}},
J_{k_{1}+r-j_{2}},\bar{J}_{k_{1}+r-j_{2}}\}$ and
the set $\{J_{k_{2}-j_{3}},\bar{J}_{k_{2}-j_{3}},
J_{k_{2}+r-j_{4}},\bar{J}_{k_{2}+r-j_{4}}\}$, since these
can only arise the cumulant expansion of 
\\*
$\Ex(J_{k_{1}-j_{1}}\bar{J}_{k_{1}-j_{1}}J_{k_{1}+r-j_{2}}\bar{J}_{k_{1}+r-j_{2}}
J_{k_{1}}\bar{J}_{k_{1}+r}\bar{J}_{k_{2}}J_{k_{2}+r}J_{k_{2}-j_{3}}\bar{J}_{k_{2}-j_{3}}
J_{k_{2}+r-j_{4}}\bar{J}_{k_{2}+r-j_{4}})$. 
Now by a careful analysis of all cumulants involving elements from both 
these two sets we observe that the largest cumulant terms are
$\cum(J_{k_{1}-j_{1}},J_{k_{2}-j_{3}})$ and 
$\cum(\bar{J}_{k_{1}-j_{1}},\bar{J}_{k_{2}-j_{3}})$ (the rest are of a lower order).  Therefore 
recalling 
that $\sum_{j_{1},j_{2},j_{3},j_{4}}w_{j_{1}}w_{j_{2}}w_{j_{3}}w_{j_{4}} = 
\sum_{j_{1},j_{2},j_{3},j_{4}=1}^{bT}\frac{1}{(bT)^{4}}\prod_{j} K(\frac{\omega_{j}}{b})$ and using 
Corollary \ref{cor:boundcumulants} gives 
\begin{eqnarray*}
&&\Ex\bigg(J_{k_{1}}\bar{J}_{k_{1}+r}\bar{J}_{k_{2}}J_{k_{2}+r}\big\{\hat{f}_{k_{1}}\hat{f}_{k_{1}+r}
- \Ex(\hat{f}_{k_{1}}\hat{f}_{k_{1}+r})\big\}\big\{\hat{f}_{k_{2}}\hat{f}_{k_{2}+r}
- \Ex(\hat{f}_{k_{2}}\hat{f}_{k_{2}+r})\big\}\bigg) \\
&\leq&  \sum_{j_{1},j_{2},j_{3},j_{4}=1}^{bT}\frac{1}{(bT)^{4}}
\prod_{j_{i}} K(\frac{\omega_{j_{i}}}{b})
\big(\frac{1}{|k_{1}+k_{2}-j_{1}-j_{3}|^{2}} + \frac{\log T}{T} \big)^{2} \\
&\leq& \frac{C}{bT}\bigg(
\frac{1}{|k_{1}-k_{2}|^{2}} + \frac{\log T}{T}\bigg),
\end{eqnarray*}
where $C$ is a finite constant independent of $k_{1}$ and $k_{2}$. 

The proof of (ii) is similar to the proof of (i), 
hence we omit the details. 
We note that if we were to show asymptotic normality of $\widehat{f}_{k}\widehat{f}_{k+r}$,
then (ii) would immediately follow from this. 

We now prove (iii). By definition of $\hat{f}_{k}\hat{f}_{k+r}$, using 
Lemmas \ref{lemma:pap} and \ref{cor:boundcumulants}, under 
Assumption \ref{assum:stat}(i) we have 
\begin{eqnarray*}
\Ex\big\{\hat{f}_{k}\hat{f}_{k+r}\big\} - f_{k}f_{k+r} &=& 
\sum_{j_{1},j_{2}}\frac{1}{(bT)^{2}}K(\frac{w_{j_{1}}}{b})K(\frac{w_{j_{2}}}{b})
\Ex\big\{J_{k-j_{1}}\bar{J}_{k-j_{1}}J_{k+r-j_{2}}\bar{J}_{k+r-j_{2}}\} - 
f(\omega_{k})f(\omega_{k+r}) \\
&=&  \sum_{j_{1}}\frac{1}{bT}K(\frac{w_{j_{1}}}{b})f_{2}(\omega_{k-j_{1}}) 
\sum_{j_{2}}\frac{1}{bT}K(\frac{w_{j_{2}}}{b})
f_{2}(\omega_{k+r-j_{1}}) - f(\omega_{k})f(\omega_{k+r}) + O(\frac{1}{bT}) \\
&=& O(b^{} + \frac{1}{bT}).
\end{eqnarray*}
Using a similar proof we can show that under Assumption \ref{assum:stat}(iv-a)
or Assumption \ref{assum:nonstat} we have 
$\Ex\big\{\hat{f}_{k}\hat{f}_{k+r}\big\} - f_{k}f_{k+r} = 
O(b^{2} + \frac{1}{bT})$. 
Thus proving (iii). 

The proof of (iv) uses Lemma \ref{lemma:pap} and 
Corollary \ref{cor:boundcumulants} and is straightforward, 
hence we omit the details. \hfill $\Box$

\vspace{3mm}

In the lemma below we prove Theorems \ref{lemma:cov-diff} and 
\ref{lemma:cov-diff-nonstat}. 

\begin{lemma}\label{lemma:diff-both}
Suppose Assumption \ref{assum:stat} holds. Then we have 
\begin{eqnarray}
\label{eq:thm3.1a}
\sqrt{T}|\widehat{c}_{T}(r) -  \widetilde{c}_{T}(r)| =
O\bigg(\frac{1}{\sqrt{bT}} + (b+\frac{1}{bT}) + 
\big(\frac{1}{bT^{1/2}} + b^{2}T^{1/2}\big)
\big(\frac{1}{|r|} + \frac{1}{T^{1/2}}\big)
\bigg). 
\end{eqnarray}
Suppose Assumption \ref{assum:nonstat} holds. Then we have
\begin{eqnarray}
\label{eq:thm4.1b}
\sqrt{T}|\widehat{c}_{T}(r) -  \widetilde{c}_{T}(r)| =
O\bigg(\frac{1}{\sqrt{bT}} + \big(\frac{1}{bT^{1/2}} + b^{2}T^{1/2}\big)
\big(\frac{1}{|r|} + \frac{1}{T^{1/2}}\big)
\bigg). 
\end{eqnarray}
\end{lemma}
PROOF. The proofs of (\ref{eq:thm3.1a}) and (\ref{eq:thm4.1b}) are very 
similar. 
Most of the time we will be obtaining the bounds under
Assumption \ref{assum:nonstat}, 
however in a few places the bounds under Assumption \ref{assum:stat} can be 
better than those under Assumption \ref{assum:nonstat}. In this case we 
will obtain the bounds under each of the Assumptions (separately). 

To prove both (\ref{eq:thm3.1a}) and (\ref{eq:thm4.1b}) we first note 
that by the mean value theorem evaluated to the second order we have 
$x^{-1/2} - y^{-1/2} = (-1/2)x^{-3/2}(x-y) + 
(1/2)(-1/2)(-3/2)x_{y}^{-5/2}(x-y)^{2}$, where
$x_{y}$ lies between $x$ and $y$. Applying this 
to the difference $\widehat{c}_{T}(r) -  \widetilde{c}_{T}(r)$
we have the expansion
\begin{eqnarray*}
\sqrt{T}|\widehat{c}_{T}(r) -  \widetilde{c}_{T}(r)| \leq \frac{1}{2}I +  \frac{3}{8}II,
\end{eqnarray*}
where 
\begin{eqnarray*}
I = \frac{1}{\sqrt{T}}\sum_{k}\frac{J_{k}\bar{J}_{k+r}}{(f_{k}f_{k+r})^{3/2}}
\big\{\hat{f}_{k}\hat{f}_{k+r} - f_{k}f_{k+r} \big\} \textrm{ and }
II = \frac{1}{\sqrt{T}}\sum_{k}\frac{J_{k}\bar{J}_{k+r}}{(\bar{f}_{k}\bar{f}_{k+r})^{5/2}}
\big\{\hat{f}_{k}\hat{f}_{k+r} - f_{k}f_{k+r} \big\}^{2}.  
\end{eqnarray*}
We consider the terms $I$ and $II$ separately. We first obtain a bound for 
$\Ex|I^{2}|$. Observe that 
$\Ex|I^{2}|= 3\big( \Ex|I_{1}^{2}| + \Ex|I_{2}^{2}|\big)$, where
\begin{eqnarray*}
I_{1} &=& \frac{1}{\sqrt{T}}\sum_{k}\frac{J_{k}\bar{J}_{k+r}}{(f_{k}f_{k+r})^{3/2}}
\big\{\hat{f}_{k}\hat{f}_{k+r} - \Ex(\hat{f}_{k}\hat{f}_{k+r}) \big\} \\
I_{2} &=&  \frac{1}{\sqrt{T}}
\sum_{k}\frac{J_{k}\bar{J}_{k+r}}{(f_{k}f_{k+r})^{3/2}}
\big\{\Ex(\hat{f}_{k}\hat{f}_{k+r}) - f_{k}f_{k+r} \big\}, 
\end{eqnarray*}
hence we will obtain the bounds $\Ex(I_{1}^{2})$ and $\Ex(I_{2}^{2})$. 
Expanding $\Ex(I_{1}^{2})$ and using Lemma \ref{lemma:difference}(i) and that
$f_{k_{1}}f_{k_{1}+r}$ is bounded away from zero gives 
\begin{eqnarray}
\label{eq:I1}
\Ex(I_{1}^{2}) \leq
\frac{1}{T}\sum_{k_{1},k_{2}}\frac{1}{(f_{k_{1}}f_{k_{1}+r}
f_{k_{2}}f_{k_{2}+r})^{3/2}}
\frac{C}{bT}\big\{\frac{1}{|k_{1}-k_{2}|^{2}} + \frac{\log T}{T}\big\}
 = O(\frac{1}{bT}). 
\end{eqnarray}
Expanding $\Ex|I_{2}^{2}|$ gives
\begin{eqnarray}
\label{eq:I2x}
\Ex(I_{2}^{2}) &\leq& \frac{C}{T}\sum_{k_{1},k_{2}}\frac{1}{(f_{k_{1}}f_{k_{1}+r}f_{k_{2}}f_{k_{2}+r})^{3/2}}
\Ex(J_{k_{1}}\bar{J}_{k_{1}+r}\bar{J}_{k_{2}}J_{k_{2}+r})\times \nonumber\\
 &&\bigg(
\Ex\big\{\hat{f}_{k_{1}}\hat{f}_{k_{1}+r}\big\} - f_{k_{1}}f_{k_{1}+r}\bigg)\bigg(
\Ex\big\{\hat{f}_{k_{2}}\hat{f}_{k_{2}+r}\big\} - f_{k_{2}}f_{k_{2}+r}\bigg). 
\end{eqnarray}
The bounds for $\Ex|I_{2}^{2}|$ differ slightly, depending on the  
assumption. 
Under Assumption \ref{assum:stat}, by using \citeA{b:bri-81}, Theorem 4.3.2,
it can be shown that 
$\Ex(J_{k_{1}}\bar{J}_{k_{1}+r}\bar{J}_{k_{2}}J_{k_{2}+r}) = O(T^{-1})$
(since $r\neq 0$). Moreover, by 
using  Lemma \ref{lemma:difference}(iii) 
we have $\Ex\big\{\hat{f}_{k_{1}}\hat{f}_{k_{1}+r}\big\} - 
f_{k_{1}}f_{k_{1}+r} = O(b+(bT)^{-1})$.
Therefore under Assumption \ref{assum:stat} we have 
\begin{eqnarray}
\label{eq:I2stat}
\Ex(I_{2}^{2}) = O\big((b + \frac{1}{bT})^{2}\big).
\end{eqnarray}
Therefore, under Assumption \ref{assum:stat}, using 
(\ref{eq:I1}) and (\ref{eq:I2stat}) gives $\Ex|I|^{2}= O(\frac{1}{bT} + 
(b + \frac{1}{bT})^{2})$ and 
\begin{eqnarray}
\label{eq:Ibdstat}
I = O_{p}(b^{} + \frac{1}{bT} + \frac{1}{\sqrt{bT}}). 
\end{eqnarray}

On the other hand, under Assumption \ref{assum:nonstat} we do not have that 
$\Ex(J_{k_{1}}\bar{J}_{k_{1}+r}\bar{J}_{k_{2}}J_{k_{2}+r}) = O(T^{-1})$, 
instead we substitute Lemma \ref{lemma:difference}(iv) into (\ref{eq:I2x})
and obtain 
\begin{eqnarray}
\label{eq:I2nonstat}
\Ex(I_{2}^{2}) &=& O\big(\{\frac{T}{r^{2}}+1\}(b^{2} + \frac{1}{bT})^{2}\big). 
\end{eqnarray}
Therefore, under Assumption \ref{assum:nonstat}, using 
(\ref{eq:I1}) and (\ref{eq:I2nonstat}) gives $\Ex|I|^{2}= O(\frac{1}{bT} + 
(\frac{T}{r^{2}} + 1)(b^{2} + \frac{1}{bT})^{2})$ and 
\begin{eqnarray}
\label{eq:I1bdnonstat}
I = O_{p}\bigg(\frac{1}{\sqrt{bT}} + \big(\frac{\sqrt{T}}{|r|} + 1\big)(b^{2} + \frac{1}{bT})\bigg). 
\end{eqnarray}
We now obtain a bound for $II$. Since the spectral density $f(\omega)$ is 
bounded away from 
zero and $\sup_{\omega}|\widehat{f}_{T}(\omega) - f(\omega)|\Pcon 0$ 
(see \cite{p:pap-09}, Lemma 6.1(ii)), we have 
$II = O_{p}(\tilde{II})$, where
\begin{eqnarray*}
\tilde{II} = \frac{1}{\sqrt{T}}\sum_{k}
\big(J_{k}\bar{J}_{k+r}
\big\{\hat{f}_{k}\hat{f}_{k+r} - f_{k}f_{k+r} \big\}^{2}\big). 
\end{eqnarray*}
To obtain a bound for $\tilde{II}$ we use that $|\tilde{II}| \leq 
3\tilde{II}_{1} +  3\tilde{II}_{2}$, where 
\begin{eqnarray*}
\tilde{II}_{1} &=& \frac{1}{\sqrt{T}}\sum_{k}
\big(|J_{k}\bar{J}_{k+r}|\big\{\hat{f}_{k}\hat{f}_{k+r} - \Ex(\hat{f}_{k}\hat{f}_{k+r}) \big\}^{2}\big) 
\textrm{ and }
\tilde{II}_{2}  = \frac{1}{\sqrt{T}}\sum_{k}
\big(|J_{k}\bar{J}_{k+r}|\big\{\Ex(\hat{f}_{k}\hat{f}_{k+r}) - f_{k}f_{k+r}  \big\}^{2}\big). 
\end{eqnarray*}
Using Cauchy-Schwarz inequality, Lemma \ref{lemma:difference}(ii,iv) we have 
\begin{eqnarray*}
\Ex|\tilde{II}_{1}| &\leq&  \frac{1}{\sqrt{T}}\sum_{k}\Ex\big(|J_{k}\bar{J}_{k+r}|^{2}\big)^{1/2}
\big(\Ex\big\{\hat{f}_{k}\hat{f}_{k+r} - \Ex(\hat{f}_{k}\hat{f}_{k+r}) \big\}^{4}\big)^{1/2} 
            = O\big(\frac{T^{1/2}}{bT}\big(\frac{1}{|r|} + \frac{1}{T^{1/2}}\big)\big). 
\end{eqnarray*}
We  now obtain a  bound for $\Ex|\tilde{II}_{2}|$. 
Using Lemma \ref{lemma:difference}(iii,iv) we have 
\begin{eqnarray*}
\Ex(\tilde{II}_{2}) &\leq& \frac{1}{\sqrt{T}}\sum_{k}
\Ex\big(|J_{k}\bar{J}_{k+r}|^{2}\big)^{1/2}
\big\{\Ex(\hat{f}_{k}\hat{f}_{k+r}) - f_{k}f_{k+r}  \big\}^{2}
= O(\sqrt{T}(b + \frac{1}{bT})^{2}\big(\frac{1}{|r|} + \frac{1}{T^{1/2}}\big)). 
\end{eqnarray*}

Therefore 
\begin{eqnarray}
\label{eq:tilde2}
\tilde{II} = 
O\bigg(\big(\frac{1}{|r|} + 
\frac{1}{T^{1/2}}\big)
\frac{1}{bT^{1/2}} + b^{2}T^{1/2} + \frac{1}{b^{2}T^{5/2}} + T^{-1/2}\bigg). 
\end{eqnarray}
Hence (\ref{eq:tilde2}) and (\ref{eq:Ibdstat}) give (\ref{eq:thm3.1a})
and (\ref{eq:tilde2}) and (\ref{eq:I1bdnonstat}) give (\ref{eq:thm4.1b}). 
\hfill $\Box$

\subsection{The variance and expectation of the covariance $\tilde{c}_{T}(r)$}\label{sec:var}

\subsubsection{Under the null}

It is straightforward to show, under Assumption \ref{assum:stat}, that
$\Ex(\sqrt{T}\tilde{c}_{T}(r)) = O(T^{-1/2})$. 
We now obtain the asymptotic variance of the $\tilde{c}_{T}(r)$ under the 
null of stationarity.  We mention that the following two lemmas
apply to nonlinear time series too. 

The following lemma immediately follows from \cite{b:bri-81}, Theorem 4.3.2. We use this result
to obtain the asymptotic variance of $\widetilde{c}_{T}(r)$, below. 
\begin{lemma}\label{lemma:statcov-J}
Let $\{X_{t}\}$ be a stationary time series where we denote the second and 
fourth order cumulants as $\kappa_{2}$ and $\kappa_{4}$. Suppose 
$\sum_{k}(1+|k|)|\kappa_{2}(k)|<\infty$ and
$\sum_{k_{1},k_{2},k_{3}}(1+|k_{i}|)|\kappa_{4}(k_{1},k_{2},k_{3})|<\infty$.
Then we have 
\begin{eqnarray}
\label{eq:Jkka}
 \cov(J_{k_{1}}J_{k_{2}},J_{k_{3}}J_{k_{4}}) &=&
\bigg(\frac{f(\omega_{k_{1}})}{T}\sum_{t=1}^{T}e^{it\omega_{k_{1}-k_{3}}} + O(\frac{1}{T})\bigg)
\bigg(\frac{f(\omega_{k_{2}})}{T}\sum_{t=1}^{T}e^{it\omega_{k_{2}-k_{4}}} + O(\frac{1}{T})\bigg)\nonumber\\
&& + \bigg(\frac{f(\omega_{k_{1}})}{T}\sum_{t=1}^{T}e^{it\omega_{k_{1}-k_{4}}} + O(\frac{1}{T})\bigg)
\bigg(\frac{f(\omega_{k_{2}})}{T}\sum_{t=1}^{T}e^{it\omega_{k_{2}-k_{3}}} + O(\frac{1}{T})\bigg)\nonumber\\
&& + (2\pi)f_{4}(\omega_{k_{1}},\omega_{k_{2}},-\omega_{k_{3}})\frac{1}{T^{2}}\sum_{t=1}^{T}e^{it\omega_{k_{1}+k_{2}-k_{3}-k_{4}}}
 + O(\frac{1}{T^{2}}). 
\end{eqnarray}
\end{lemma}

\begin{lemma}\label{lemma:stat-var-general}
Suppose the assumptions in Lemma \ref{lemma:statcov-J} hold.
Then we have 
\begin{eqnarray}
\label{eq:summandcovariance}
&& \cov\big(\sqrt{T} \Re\widetilde{c}_{T}(r_{1}), \sqrt{T}\Re \widetilde{c}_{T}(r_{2})\big)
 = \cov\big( \sqrt{T}\Im \widetilde{c}_{T}(r_{1}), \sqrt{T}\Im \widetilde{c}_{T}(r_{2})\big) \nonumber\\
 &=& 
\left\{
\begin{array}{cc}
 O(T^{-1}) & r_{1}\neq r_{2} \\ 
1 + \frac{1}{2}
 \frac{2\pi}{2T^{2}}\sum_{k_{1},k_{2}=1}^{T}g_{T,k_{1}}^{(r)}g_{T,k_{2}}^{(r)}
f_{4}(\omega_{k_{1}},-\omega_{k_{1}+r},-\omega_{k_{2}}) + O(\frac{1}{T}) & r_{1} = r_{2} = r\\
\end{array}
\right.
\end{eqnarray}
Furthermore if the tri-spectra $f_{4}(\omega_{1},\omega_{2},\omega_{3})$ 
is Lipschitz continuous we have 
\begin{eqnarray}
\label{eq:integralcovariance}
&& \cov\big( \sqrt{T} \Re\widetilde{c}_{T}(r_{1}), \sqrt{T}\Re \widetilde{c}_{T}(r_{2})\big)
 = \cov\big(\sqrt{T} \Im \widetilde{c}_{T}(r_{1}), \sqrt{T}\Im \widetilde{c}_{T}(r_{2})\big) \nonumber\\
 &=& 
\left\{
\begin{array}{cc}
 O(T^{-1}) & r_{1}\neq r_{2} \\ 
1 + \frac{1}{4\pi}\int_{0}^{2\pi} \int_{0}^{2\pi}\frac{f_{4}(\omega_{1},-\omega_{1}-\omega_{r},-\omega_{2})}{
\sqrt{f(\omega_{1})f(\omega_{1}+\omega_{r})f(\omega_{2})f(\omega_{2}+\omega_{r})}}d\omega_{1}d\omega_{2}
 +O(\frac{1}{T}) & r_{1} = r_{2} = r\\
\end{array}
\right.
\end{eqnarray}
and for all $r_{1},r_{2}$, 
$\cov(\sqrt{T}\Re \widetilde{c}_{T}(r_{1}), \sqrt{T}\Im \widetilde{c}_{T}(r_{2})) = O(\frac{1}{T})$. 
\end{lemma}
PROOF. To prove the result we use that $\Re \widetilde{c}_{T}(r) = 
\frac{1}{2}(\widetilde{c}_{T}(r)+\overline{\widetilde{c}}_{T}(r))$ and 
$\Im \widetilde{c}_{T}(r) = 
\frac{-i}{2}(\widetilde{c}_{T}(r)+\overline{\widetilde{c}}_{T}(r))$, and 
evaluate $\cov(\sqrt{T}\tilde{c}_{T}(r_{1}),\sqrt{T}\tilde{c}_{T}(r_{2}))$, 
$\cov(\sqrt{T}\tilde{c}_{T}(r_{1}),\sqrt{T}\overline{\tilde{c}}_{T}(r_{2}))$
and $\cov(\sqrt{T}\overline{\tilde{c}}_{T}(r_{1}),
\sqrt{T}\overline{\tilde{c}}_{T}(r_{2}))$.
Expanding $\cov(\sqrt{T}\tilde{c}_{T}(r_{1}),\sqrt{T}\tilde{c}_{T}(r_{2}))$  
gives 
\begin{eqnarray*}
\cov(\sqrt{T}\tilde{c}_{T}(r_{1}),\sqrt{T}\tilde{c}_{T}(r_{2})) &=& 
\frac{1}{T}\sum_{k_{1},k_{2}=1}^{T}
g_{T,k_{1}}^{(r_{1})}
g_{T,k_{2}}^{(r_{2})}\cov(J_{k_{1}}\overline{J}_{k_{1}+r_{1}},
J_{k_{2}}\overline{J}_{k_{2}+r_{2}}), 
\end{eqnarray*}
where $g_{T,k}^{(r)} = f(\omega_{k})^{-1/2}f(\omega_{k}+\omega_{r})^{-1/2}$. 
Substituting (\ref{eq:Jkka}) into the above it is easy to show that for 
$r_{1}\neq r_{2}$ we have 
$\cov(\sqrt{T}\tilde{c}_{T}(r_{1}),\sqrt{T}\tilde{c}_{T}(r_{2})) = O(T^{-1})$ and for
$r:=r_{1}=r_{2}$ we have 
\begin{eqnarray*}
&& \cov(\sqrt{T}\tilde{c}_{T}(r),\sqrt{T}\tilde{c}_{T}(r)) \\
&=& 
\frac{2}{T}\sum_{k=1}^{T}(g_{T,k}^{(r)})^{2}
f(\omega_{k})f(\omega_{k+r}) 
+\frac{1}{T^{2}}\sum_{k_{1},k_{2}=1}^{T}g_{T,k_{1}}^{(r)}g_{T,k_{2}}^{(r)}
  f_{4}(\omega_{k_{1}},-\omega_{k_{1}+r},-\omega_{k_{2}}) + 
 O(\frac{1}{T})  \\
&=&
2 + \frac{2\pi}{T^{2}}\sum_{k_{1},k_{2}=1}^{T}g_{T,k_{1}}^{(r)}g_{T,k_{2}}^{(r)}
f_{4}(\omega_{k_{1}},-\omega_{k_{1}+r},-\omega_{k_{2}}) + O(\frac{1}{T}). 
\end{eqnarray*}
The same method gives us a similar bound for 
$\cov(\sqrt{T}\overline{\tilde{c}_{T}(r)},
\sqrt{T}\overline{\tilde{c}_{T}(r)})$. 
Similarly it can be shown that unless $r_{2} = T-r_{1}$ we have 
$\cov(\sqrt{T}\tilde{c}_{T}(r_{1}),\overline{\sqrt{T}\tilde{c}_{T}(r_{2})}) =
O(T^{-1})$. Also, for $r_{1}\neq r_{2}$ we have 
$\cov(\overline{\sqrt{T}\tilde{c}_{T}(r_{1})},\overline{\sqrt{T}\tilde{c}_{T}(r_{2})}) = O(T^{-1})$
Altogether this gives us (\ref{eq:summandcovariance}). 

We now obtain (\ref{eq:integralcovariance}) by using 
(\ref{eq:summandcovariance}) and replacing the sum with an integral. 
By using that the spectral density and tri-spectra $f(\cdot)$ and 
$f_{4}(\cdot)$ are Lipschitz continuous we can replace the summand 
above with an integral to give 
\begin{eqnarray*}
\cov(\sqrt{T}\tilde{c}_{T}(r),\sqrt{T}\tilde{c}_{T}(r)) 
 &=& 2 + \frac{1}{2\pi}\int_{0}^{\pi} \int_{0}^{\pi}\frac{f_{4}(\omega_{1},-\omega_{1}-\omega_{r},-\omega_{2})}{
\sqrt{f(\omega_{1})f(\omega_{1}+\omega_{r})f(\omega_{2})f(\omega_{2}+\omega_{r})}}d\omega_{1}d\omega_{2}
+ O(\frac{1}{T}). 
\end{eqnarray*}
A similar result can be obtained for 
$\cov(\sqrt{T}\overline{\tilde{c}_{T}(r)},
\sqrt{T}\overline{\tilde{c}_{T}(r)})$. 
Substituting the above into (\ref{eq:summandcovariance}) gives
(\ref{eq:integralcovariance}). 
\hfill $\Box$

\vspace{3mm}

\begin{lemma}\label{lemma:phase-cont-stat}
Suppose Assumption \ref{assum:stat} holds. Then the spectral density $f(\omega)$ and  
the phase function $\phi(\omega)$, 
satisfy $\sup_{\omega}|f^{\prime}(\omega)|<\infty$ and 
$\sup_{\omega}|\phi^{\prime}(\omega)| < \infty$. 
\end{lemma}
PROOF. The fact that $\sup_{\omega}|f^{\prime}(\omega)|<\infty$ follows 
immediately from 
Assumption \ref{assum:stat}(i). To prove that 
$\sup_{\omega}|\phi^{\prime}(\omega)| < \infty$
we recall that $\phi(\omega) = \arctan \frac{\Im A(\omega)}{\Re A(\omega)}$. 
Differentiating $\phi(\omega)$ with respect to $\omega$ and using the 
chain rule gives
\begin{eqnarray*}
\frac{d\phi(\omega)}{d\omega} = \frac{1}{1+\big(  
\frac{\Im A(\omega)}{\Re A(\omega)} \big)^{2}}
\times \frac{1}{[\Re A(\omega)]^{2}}
\bigg(\Im A(\omega)\frac{d \Re A(\omega)}{d\omega} -
\Re A(\omega)\frac{d \Im A(\omega)}{d\omega} \bigg). 
\end{eqnarray*}
Under Assumption \ref{assum:stat}(i,iii) we have 
$\inf_{\omega}\Re A(\omega) > 0$, $\sup_{\omega}|A(\omega)| < \infty$ and 
$\sup_{\omega}|\frac{d A(\omega)}{d\omega }| < \infty$. Therefore 
$\sup_{\omega} |\frac{d\phi(\omega)}{d\omega}|<\infty$. 
Thus giving the required result. \hfill $\Box$

\vspace{3mm}

{\bf PROOF of Lemma \ref{lemma:variance-stat}}
We use Lemma \ref{lemma:stat-var-general} to prove the result. 
We note that Assumption \ref{assum:stat} satisfy 
the assumptions in Lemma \ref{lemma:stat-var-general}. 
Therefore we have the identity in (\ref{eq:summandcovariance}). 
However in the case that the time series is linear this expression can be 
simplified. 
It is well know that for a linear time series the tri-spectra can be written in terms of 
the transfer function $A(\omega)$ that is 
\begin{eqnarray*}
f_{4}(\omega_{1},\omega_{2},\omega_{3}) = 
\frac{\kappa_{4}}{2\pi}A(\omega_{1})A(\omega_{2})A(\omega_{3})
A(-\omega_{1}-\omega_{2}-\omega_{3}). 
\end{eqnarray*}
Now we recall that for a linear time series $A(\omega) = \sqrt{f(\omega)}\exp(i\phi(\omega))$ hence 
substituting this into the ratio in (\ref{eq:summandcovariance}) gives
\begin{eqnarray*}
2\pi\frac{f_{4}(\omega_{k_{1}},-\omega_{k_{2}}-\omega_{r},-\omega_{k_{2}})}{
\sqrt{f(\omega_{k_{1}})f(\omega_{k_{1}}+\omega_{2})f(\omega_{k_{2}})f(\omega_{k_{2}}+\omega_{r})}} 
= \kappa_{4}
\exp\big(i[\phi(\omega_{k_{1}}) - \phi(\omega_{k_{1}}+\omega_{r}) - \phi(\omega_{k_{2}}) + 
\phi(\omega_{k_{2}}+\omega_{r})]
\big). 
\end{eqnarray*}
Substituting the above into (\ref{eq:summandcovariance}) gives
\begin{eqnarray*}
\cov(\widetilde{c}_{T}(r_{}),\widetilde{c}_{T}(r_{})) = 
1 + \frac{\kappa_{4}}{2}\bigg|\frac{1}{T}\sum_{k}\exp(i[\phi_{k}-\phi _{k+r_{}}])\bigg|^{2}.
\end{eqnarray*}
Finally we note that due to Lemma \ref{lemma:phase-cont-stat}, the phase $\phi(\cdot)$ is Lipschitz
continuous, hence $\exp(i\phi(\omega))$ is Lipschitz continuous, and we can replace
the summand in the above with an integral to give 
\begin{eqnarray*}
\cov(\widetilde{c}_{T}(r_{}),\widetilde{c}_{T}(r_{})) = 
\sigma^{2} + \frac{\kappa_{4}}{2}\bigg|\frac{1}{2\pi}
\int_{0}^{2\pi}\exp(i[\phi(\omega)-
\phi(\omega+\frac{2\pi r_{}}{T})])d\omega\bigg|^{2} + O(\frac{1}{T}). 
\end{eqnarray*}
This completes the proof. \hfill $\Box$

\subsubsection{The alternative of local stationarity}

We now consider some of the moment properties of $\widetilde{c}_{T}(r)$ under
the assumption of local stationarity. 

\begin{lemma}\label{lemma:KJvar}
Suppose Assumption \ref{assum:nonstat} holds. Then we have 
\begin{eqnarray}
\label{eq:Jkk}
\cov(J_{k_{1}}J_{k_{2}},J_{k_{3}}J_{k_{4}}) &=&
\big\{F_{2}(k_{1}-k_{3};\omega_{k_{1}})F_{2}(k_{2}-k_{4};\omega_{k_{2}}) + 
  F_{2}(k_{1}-k_{4};\omega_{k_{1}})F_{2}(k_{2}-k_{3};\omega_{k_{2}})\big\} \nonumber\\
&+& \frac{(2\pi)}{T}F_{4}(k_{1}+k_{2}-k_{3}-k_{3};\omega_{k_{1}},\omega_{k_{2}},-\omega_{k_{3}}) \nonumber\\
&+&  O\big(\frac{(\log T)^{3}}{T^{2}} + \frac{\log T}{T}\big(F_{2}(k_{1}-k_{3};\omega_{k_{1}}) + 
F_{2}(k_{2}-k_{4};\omega_{k_{2}})+ \nonumber\\
&& F_{2}(k_{1}-k_{4};\omega_{k_{1}}) +
F_{2}(k_{2}-k_{3};\omega_{k_{2}})\big)\big),  
\end{eqnarray}
where $\{F_{2}(\cdot;\omega)\}$ and 
$\{F_{4}(\cdot;\omega_{k_{1}},\omega_{k_{2}},\omega_{k_{3}})\}$ are
defined in (\ref{eq:FTomega}).
\end{lemma}
PROOF. Expanding $\cov(J_{k_{1}}J_{k_{2}},J_{k_{3}}J_{k_{4}})$ in terms of cumulants gives
\begin{eqnarray*}
\cov(J_{k_{1}}J_{k_{2}},J_{k_{3}}J_{k_{4}}) = 
\cov(J_{k_{1}},\bar{J}_{k_{3}})\cov(J_{k_{2}},\bar{J}_{k_{4}}) 
 +  \cov(J_{k_{1}},\bar{J}_{k_{4}})\cov(J_{k_{2}},\bar{J}_{k_{3}}) + 
\cum(J_{k_{1}},J_{k_{2}},\bar{J}_{k_{3}},\bar{J}_{k_{4}}), 
\end{eqnarray*}
finally substituting Corollary \ref{cor:boundcumulants} into the above gives 
the result. \hfill $\Box$

\vspace{3mm}

{\bf PROOF of Lemma \ref{lemma:meanalt} equations  
(\ref{eq:integrated-spec}) and (\ref{eq:bias-LS})} 
We first prove (\ref{eq:integrated-spec}). We note that
from the definition of $\widehat{f}_{T}(\omega_{k})$ in (\ref{eq:spectral-estimator}) and 
under Assumption \ref{assum:nonstat} we have 
\begin{eqnarray}
\label{eq:hatbias}
\big|\Ex(\widehat{f}_{T}(\omega_{k}))  -  f(\omega_{k}) \big| = 
\big|\sum_{j}\frac{1}{bT}K(\frac{\omega_{k}-\omega_{j}}{b})\Ex(|J_{T}(\omega_{j})|^{2}) 
- f(\omega_{k}) \big| = O(b^{2}). 
\end{eqnarray}
We now obtain $\var(\widehat{f}_{T}(\omega_{k}))$. We observe that
\begin{eqnarray*}
\var(\widehat{f}_{T}(\omega_{k})) = \sum_{j_{1},j_{2}}\frac{1}{(bT)^{2}}K(\frac{\omega_{k}-\omega_{j_{1}}}{b})
K(\frac{\omega_{k}-\omega_{j_{2}}}{b})\cov(|J_{T}(\omega_{j_{1}})|^{2},|J_{T}(\omega_{j_{2}})|^{2}).
\end{eqnarray*}
Now we substitute (\ref{eq:Jkk}) into the above to give 
\begin{eqnarray*}
\var(\widehat{f}_{T}(\omega_{k})) 
&\leq& C\sum_{j_{1},j_{2}}\frac{1}{(bT)^{2}}K(\frac{\omega_{k}-\omega_{j_{1}}}{b})
K(\frac{\omega_{k}-\omega_{j_{2}}}{b})
\{|F_{2}(j_{1}-j_{2};\omega_{j_{1}})F_{2}
(j_{2}-j_{1};\omega_{j_{1}})| + \nonumber\\
&&  |F_{2}(j_{2}-j_{1};\omega_{j_{1}})
F_{2}(j_{1}-j_{2};-\omega_{j_{1}})|
+\frac{2\pi}{T}
F_{4}(0;\omega_{j_{1}},-\omega_{j_{1}},\omega_{j_{2}})\big\}
  + O\big(\frac{(\log T)^{3}}{T^{2}}\big). 
\end{eqnarray*}
We observe from the above that the covariance terms dominate 
the fourth order cumulant term. Moreover, by using  
Lemma \ref{lemma:fourierstuff} we have 
$\sup_{\omega}\sum_{s}|F_{2}(s;\omega)|<\infty$, which gives
$\var(\widehat{f}_{T}(\omega_{k})) = O(\frac{1}{bT})$. 
This together with (\ref{eq:hatbias}) gives (\ref{eq:integrated-spec}). 

We now prove (\ref{eq:bias-LS}). Using Lemma \ref{lemma:pap} for $n=2$ gives
\begin{eqnarray}
\label{eq:bias-nonstat}
\Ex(\tilde{c}_{T}(r)) 
 &=& \frac{1}{T}\sum_{k=1}^{T}\frac{1}{[f(\omega_{k})f(\omega_{k} + \omega_{r})]^{1/2}}
\frac{1}{T}\sum_{t=1}^{T}f(\frac{t}{T},\omega_{k})\exp(-it\omega_{r}) + O(\frac{\log T}{T}). 
\end{eqnarray}
Now by using replacing sum with integral and 
using Lemma \ref{lemma:FDFT} (noting 
\\*
$F_{2,T}(-r;\omega_{k}) = \frac{1}{T}\sum_{t=1}^{T}f(\frac{t}{T},\omega_{k})\exp(-it\omega_{r})$ and
$F_{2}(-r;\omega_{k}) = 
\int_{0}^{1}f(u,\omega_{k})\exp(-i2\pi r u )du$) gives 
\begin{eqnarray}
\label{eq:bias-nonstat}
\Ex(\tilde{c}_{T}(r)) 
 &=& \frac{1}{2\pi}\int_{0}^{1}\int_{0}^{2\pi}\frac{f(u,\omega)}{[f(\omega)f(\omega + \omega_{r})]^{1/2}}
\exp(-i2\pi u r)dud\omega + O(\frac{\log T}{T} + \frac{1}{T^{2}}), 
\end{eqnarray}
thus we have (\ref{eq:bias-LS}). \hfill $\Box$

\vspace{3mm}

To prove (\ref{eq:limitvar-nonstat}) in Lemma \ref{lemma:meanalt}, 
we evaluate the limiting variance of $\tilde{c}_{T}(t)$ under the alternative of 
local stationarity. 
\begin{lemma}\label{lemma:local-stat-cov}
Suppose Assumption \ref{assum:nonstat} holds. Then we have 
\begin{eqnarray}
\label{eq:tildevarC}
\cov(\Re\sqrt{T}\tilde{c}_{T}(r_{1}),\Re\sqrt{T}\tilde{c}_{T}(r_{2})) = 
\frac{1}{4}\big(\Gamma_{T,r_{1},r_{2}}^{(1)} + \Gamma_{T,r_{1},r_{2}}^{(2)}
+ \Gamma_{T,r_{2},r_{1}}^{(2)} + \Gamma_{T,r_{1},r_{2}}^{(3)} \big) + O(\frac{\log T}{T}),
\end{eqnarray}
\begin{eqnarray*}
\cov(\Re\sqrt{T}\tilde{c}_{T}(r_{1}),\Im\sqrt{T}\tilde{c}_{T}(r_{2})) = 
\frac{-i}{4}\big(\Gamma_{T,r_{1},r_{2}}^{(1)} + \Gamma_{T,r_{1},r_{2}}^{(2)}
- \Gamma_{T,r_{2},r_{1}}^{(2)} - \Gamma_{T,r_{1},r_{2}}^{(3)} \big) + O(\frac{\log T}{T}),
\end{eqnarray*}
\begin{eqnarray*}
\cov(\Im\sqrt{T}\tilde{c}_{T}(r_{1}),\Im\sqrt{T}\tilde{c}_{T}(r_{2})) = 
\frac{1}{4}\big(\Gamma_{T,r_{1},r_{2}}^{(1)} - \Gamma_{T,r_{1},r_{2}}^{(2)}
- \Gamma_{T,r_{2},r_{1}}^{(2)} + \Gamma_{T,r_{1},r_{2}}^{(3)} \big) + O(\frac{\log T}{T}),
\end{eqnarray*}
where 
\begin{eqnarray*}
&&\Gamma_{T,r_{1},r_{2}}^{(1)} = \frac{1}{T}\sum_{k_{1},k_{2}}g_{T,k_{1}}^{(r_{1})}g_{T,k_{2}}^{(r_{2})}
\big\{
F_{2}(k_{1}-k_{2};\omega_{k_{1}})
F_{2}(-k_{1}-r_{1} +k_{2}+r_{2};-\omega_{k_{1}+r_{1}}) + \nonumber\\
&& F_{2}(k_{1}+k_{2}+r_{2};\omega_{k_{1}})
F_{2}(-k_{1}-r_{1}-k_{2};-\omega_{k_{1}+r_{1}})\big\}  +
 \frac{1}{T^{2}}\sum_{k_{1},k_{2}} g_{T,k_{1}}^{(r_{1})}g_{T,k_{2}}^{(r_{2})}
F_{4}
(r_{2}-r_{1};\omega_{k_{1}},-\omega_{k_{1}+r_{1}},-\omega_{k_{2}}),  
\end{eqnarray*}
\begin{eqnarray*}
&&\Gamma_{T,r_{1},r_{2}}^{(2)} = \frac{1}{T}\sum_{k_{1},k_{2}}g_{T,k_{1}}^{(r_{1})}g_{T,k_{2}}^{(r_{2})}
\big\{F_{2}(k_{1}+k_{2};\omega_{k_{1}})F_{2}(-k_{1}-r_{1}-k_{2}-r_{2};-\omega_{k_{1}+r_{1}}) + \nonumber\\
&& F_{2}(k_{1}-k_{2}-r_{2};\omega_{k_{1}})F_{2}(-k_{1}-r_{1}+k_{2};-\omega_{k_{1}+r_{1}})\big\}  
 + \frac{1}{T^{2}}\sum_{k_{1},k_{2}} g_{T,k_{1}}^{(r_{1})}g_{T,k_{2}}^{(r_{2})}
F_{4}(-r_{2}-r_{1};\omega_{k_{1}},-\omega_{k_{1}+r_{1}},-\omega_{k_{2}}),  
\end{eqnarray*}
\begin{eqnarray*}
&&\Gamma_{T,r_{1},r_{2}}^{(3)} = \frac{1}{T}\sum_{k_{1},k_{2}}g_{T,k_{1}}^{(r_{1})}g_{T,k_{2}}^{(r_{2})}
\big\{F_{2}(-k_{1}+k_{2};-\omega_{k_{1}})F_{2}(k_{1}+r_{1}-k_{2}-r_{2};\omega_{k_{1}+r_{1}}) + \nonumber\\
&&  F_{2}(-k_{1}-k_{2}-r_{2};-\omega_{k_{1}})F_{2}(k_{1}+r_{1}+k_{2};\omega_{k_{1}+r_{1}})\big\} 
 + \frac{1}{T^{2}}\sum_{k_{1},k_{2}} g_{T,k_{1}}^{(r_{1})}g_{T,k_{2}}^{(r_{2})}
F_{4}(r_{1}-r_{2};-\omega_{k_{1}},\omega_{k_{1}+r_{1}},-\omega_{k_{2}}), 
\end{eqnarray*}
and the coefficients $F_{2}(\cdot)$ and $F_{4}(\cdot)$ are defined in 
(\ref{eq:FTomega}) and  
$g_{T,k}^{(r)} = 
\big\{f(\omega_{k})f(\omega_{k+r})\big\}^{-1/2}$.  
\end{lemma}
PROOF. To prove (\ref{eq:tildevarC}) we use $\Re \widetilde{c}_{T}(r) = 
\frac{1}{2}(\widetilde{c}_{T}(r)+\overline{\widetilde{c}}_{T}(r))$ and 
$\Im \widetilde{c}_{T}(r) = 
\frac{-i}{2}(\widetilde{c}_{T}(r)+\overline{\widetilde{c}}_{T}(r))$, and 
$\cov(\sqrt{T}\tilde{c}_{T}(r_{1}),\sqrt{T}\overline{\tilde{c}}_{T}(r_{2}))$
and $\cov(\sqrt{T}\overline{\tilde{c}}_{T}(r_{1}),\sqrt{T}\overline{\tilde{c}}_{T}(r_{2}))$.
Expanding $\cov(\sqrt{T}\tilde{c}_{T}(r_{1}),\sqrt{T}\tilde{c}_{T}(r_{2}))$ we have 
\begin{eqnarray*}
\cov(\sqrt{T}\tilde{c}_{T}(r_{1}),\sqrt{T}\tilde{c}_{T}(r_{2})) = 
\frac{1}{T}\sum_{k_{1},k_{2}}g_{T,k_{1}}^{(r_{1})}g_{T,k_{2}}^{(r_{2})}\cov(J_{k_{1}}\overline{J}_{k_{1}+r_{1}},
J_{k_{2}}\overline{J}_{k_{2}+r_{2}}), 
\end{eqnarray*}
now by substituting (\ref{eq:Jkk}) into the above we obtain 
\begin{eqnarray*}
\cov(\sqrt{T}\tilde{c}_{T}(r_{1}),\sqrt{T}\tilde{c}_{T}(r_{2})) = \Gamma_{T,r_{1},r_{2}}^{(1)} + 
O(\frac{\log T}{T}). 
\end{eqnarray*}
Similar results can be obtained for $\cov(\sqrt{T}\tilde{c}_{T}(r_{1}),
\sqrt{T}\overline{\tilde{c}}_{T}(r_{2}))$
and
$\cov(\sqrt{T}\overline{\tilde{c}}_{T}(r_{1}),
\sqrt{T}\overline{\tilde{c}}_{T}(r_{2}))$.
Using this we obtain the required result.  
\hfill $\Box$

\vspace{3mm}

{\bf PROOF of Lemma \ref{lemma:meanalt}, equation (\ref{eq:limitvar-nonstat})}
This immediately follows from Lemma \ref{lemma:local-stat-cov}. \hfill $\Box$

\subsection{Asymptotic normality}\label{sec:asym-normality}

In this section we prove asymptotic normality of $\sqrt{T}\tilde{c}_{T}(r)$.  
Since the locally stationary linear time series model includes the stationary 
time series model as a special case we show asymptotic normality of the more 
general locally stationary model. 
We start by approximating $\sqrt{T}\tilde{c}_{T}(r)$ with a 
random variable which 
only involves current innovations
$\{\varepsilon_{t}\}_{t=1}^{T}$. We make this approximation in 
order to use the 
martingale central limit theorem to prove  
asymptotic normality of $\sqrt{T}\tilde{c}_{T}(r)$.
In this section, we will make frequent  
appeals to Lemma \ref{lemma:GDFT}. 

Using that the locally stationary time series model  
$X_{t,T}$ satisfies (\ref{eq:tvma}) we have can write 
$\sqrt{T}\tilde{c}_{T}(r)$ as
\begin{eqnarray*}
&&\sqrt{T}\tilde{c}_{T}(r)  \\
&&= \frac{1}{T^{3/2}}
\sum_{k=1}^{T}\frac{1}{f(\omega_{k})^{1/2}f(\omega_{k} + \omega_{r})^{1/2}}
\sum_{t,\tau=1}^{T}\exp(i(t-\tau)\omega_{k})\exp(-i\tau\omega_{r})\sum_{j_{1},j_{2}=0}^{\infty}
\psi_{t,T}(j_{1})\psi_{\tau,T}(j_{2})\varepsilon_{t-j_{1}}\varepsilon_{\tau-j_{2}} \\
&=& \frac{1}{T^{1/2}}
\sum_{t,\tau=1}^{T}G_{T,\omega_{r}}(t-\tau)
\exp(-i\tau\omega_{r})\sum_{j_{1},j_{2}=0}^{\infty}
\psi_{t,T}(j_{1})\psi_{\tau,T}(j_{2})\big(\varepsilon_{t-j_{1}}\varepsilon_{\tau-j_{2}}
- \Ex(\varepsilon_{t-j_{1}}\varepsilon_{\tau-j_{2}}) \big),
\end{eqnarray*}
where $\{G_{T,\omega_{r}}(s)\}$ is the DFT
defined in (\ref{eq:fourierCOA}). 

We now partition $\sqrt{T}\tilde{c}_{T}(r)$ into terms which involve 
`present' and `past' innovation, that is 
\begin{eqnarray}
\label{eq:decom=past-pres}
\sqrt{T}\big(\tilde{c}_{T}(r) - \Ex(\tilde{c}_{T}(r)\big) = 
\sqrt{T}\big(d_{T}(r) + V_{T}(r)\big), 
\end{eqnarray}
where 
\begin{eqnarray*}
d_{T}(r) &=& \frac{1}{T}
\sum_{t,\tau=1}^{T}G_{T,\omega_{r}}(t-\tau)
\exp(-i\tau\omega_{r})\sum_{0\leq j_{1}\leq t-1}\,\sum_{0\leq j_{2}\leq \tau-1}
\psi_{t,T}(j_{1})\psi_{\tau,T}(j_{2})\big(\varepsilon_{t-j_{1}}\varepsilon_{\tau-j_{2}}
- \Ex(\varepsilon_{t-j_{1}}\varepsilon_{\tau-j_{2}}) \big) \\
V_{T}(r) &=& \frac{1}{T}
\sum_{t,\tau=1}^{T}G_{T,\omega_{r}}(t-\tau)
\exp(-i\tau\omega_{r})
\sum_{j_{1}\geq t-1 \textrm{ or }j_{2}\geq \tau-1 }
\psi_{t,T}(j_{1})\psi_{\tau,T}(j_{2})\big(\varepsilon_{t-j_{1}}\varepsilon_{\tau-j_{2}}
- \Ex(\varepsilon_{t-j_{1}}\varepsilon_{\tau-j_{2}}) \big).  
\end{eqnarray*}
In the following lemma we obtain a bound for the remainder $V_{T}(r)$. Later we
will show asymptotic normality of $d_{T}(r)$. 
\begin{lemma}\label{lemma:Rtbound}
Suppose Assumption \ref{assum:nonstat} hold. Then we have
\begin{eqnarray*}
\big(T^{1/2}\Ex|V_{T}(r)|^{2}\big)^{1/2} \leq CT^{-1/2},
\end{eqnarray*}
for some finite constant $C$.
\end{lemma}
PROOF. We first observe that $\sqrt{T}V_{T}(r) = I_{1} + I_{2} + I_{3}$, where
\begin{eqnarray*}
I_{1} &=& \frac{1}{T^{1/2}}
\sum_{t,\tau=1}^{T}G_{T,\omega_{r}}(t-\tau)
\exp(-i\tau\omega_{r})
\sum_{j_{1}\geq t-1}\sum_{0\leq j_{2}\leq \tau-1}
\psi_{t,T}(j_{1})\psi_{\tau,T}(j_{2})\big(\varepsilon_{t-j_{1}}\varepsilon_{\tau-j_{2}}
- \Ex(\varepsilon_{t-j_{1}}\varepsilon_{\tau-j_{2}}) \big) 
\end{eqnarray*}
\begin{eqnarray*}
I_{2} &=& \frac{1}{T^{1/2}}
\sum_{t,\tau=1}^{T}G_{T,\omega_{r}}(t-\tau)
\exp(-i\tau\omega_{r})
\sum_{j_{2}\geq \tau-1}\sum_{0\leq j_{1}\leq t-1 }
\psi_{t,T}(j_{1})\psi_{\tau,T}(j_{2})\big(\varepsilon_{t-j_{1}}\varepsilon_{\tau-j_{2}}
- \Ex(\varepsilon_{t-j_{1}}\varepsilon_{\tau-j_{2}}) \big) 
\end{eqnarray*}
\begin{eqnarray*}
I_{3} &=& \frac{1}{T^{1/2}}
\sum_{t,\tau=1}^{T}G_{T,\omega_{r}}(t-\tau)
\exp(-i\tau\omega_{r})
\sum_{j_{2}\geq \tau-1}\sum_{j_{1}\geq t-1}
\psi_{t,T}(j_{1})\psi_{\tau,T}(j_{2})\big(\varepsilon_{t-j_{1}}\varepsilon_{\tau-j_{2}}
- \Ex(\varepsilon_{t-j_{1}}\varepsilon_{\tau-j_{2}}) \big). 
\end{eqnarray*}
We first show that $\Ex(I_{1}^{2})^{1/2} = O(T^{-1/2})$. By the Minkowski's inequality we have 
\begin{eqnarray*}
\Ex(I_{1}^{2})^{1/2} \leq \frac{1}{T^{1/2}}
\sum_{t,\tau=1}^{T}|G_{T,\omega_{r}}(t-\tau)|
\bigg\{\Ex\bigg(\sum_{j_{1}\geq t-1}\sum_{0\leq j_{2}\leq \tau-1}
\psi_{t,T}(j_{1})\psi_{\tau,T}(j_{2})\big(\varepsilon_{t-j_{1}}\varepsilon_{\tau-j_{2}}
- \Ex(\varepsilon_{t-j_{1}}\varepsilon_{\tau-j_{2}}) \big) \bigg)^{2}\bigg\}^{1/2}. 
\end{eqnarray*}
It can be shown that 
\begin{eqnarray*}
&&\Ex\bigg(\sum_{j_{1}\geq t-1}\sum_{0\leq j_{2}\leq \tau-1}
\psi_{t,T}(j_{1})\psi_{\tau,T}(j_{2})\big(\varepsilon_{t-j_{1}}\varepsilon_{\tau-j_{2}}
- \Ex(\varepsilon_{t-j_{1}}\varepsilon_{\tau-j_{2}}) \big) \bigg)^{2} \\
&\leq& \big(\var(\varepsilon_{t})^{2}+ \var(\varepsilon_{t}^{2})\big)
\bigg[ \sum_{j_{1}\geq t}|\psi_{t,T}(j_{1})|^{2}
\bigg]\bigg[ \sum_{j_{2}=0}^{\infty}|\psi_{\tau,T}(j_{2})|^{2}
\bigg]. 
\end{eqnarray*}
Substituting this into the bound for $\Ex(I_{1}^{2})$, under 
Assumption \ref{assum:nonstat}
and using Lemma \ref{lemma:GDFT} we have
\begin{eqnarray*}
\Ex(I_{1}^{2})^{1/2} &\leq & \frac{1}{T^{1/2}}\sup_{\tau}
\bigg[ \sum_{j_{2}=0}^{\infty}|\psi_{\tau,T}(j_{2})|^{2}
\bigg]^{1/2}
\bigg[\sum_{s=1}^{T}|G_{T,\omega_{r}}(s)|\bigg]
\sum_{t=1}^{T}\bigg[ \sum_{j_{1}\geq t}|\psi_{t,T}(j_{1})|^{2}
\bigg]^{1/2}  \\
&& \frac{1}{T^{1/2}}\sup_{\tau}
\bigg[ \sum_{j_{2}=0}^{\infty}|\psi_{\tau,T}(j_{2})|^{2}
\bigg]^{1/2}
\bigg[\sum_{s=1}^{T}|G_{T,\omega_{r}}(s)|\bigg]
\sum_{t=1}^{T}\sum_{j_{1}\geq t}|\psi_{t,T}(j_{1})| = O(T^{-1/2}). 
\end{eqnarray*}
Using a similar method we can show that $\Ex(I_{2}^{2})^{1/2} = 
O(T^{-1/2})$ and 
$\Ex(I_{3}^{2})^{1/2} = O(T^{-1/2})$. Thus we obtain the 
result. \hfill $\Box$

\vspace{3mm}

Therefore the above lemma shows that 
$\sqrt{T}\big(\tilde{c}_{T}(r) - \Ex(\tilde{c}_{T}(r)\big) = 
\sqrt{T}d_{T}(r) + O_{p}(T^{-1/2})$.

\begin{remark}
Now it is worth noting that in the case that $\{X_{t}\}$ is a stationary
linear time series, then $d_{T}(r)$ has an interesting form. That 
is, it is straightforward to show (using \cite{b:pri-88}, Theorem 
6.2.1) that 
\begin{eqnarray*}
\sqrt{T}\tilde{c}_{T}(r) 
 &=&\frac{1}{\sqrt{T}}\sum_{k=1}^{T}J_{\varepsilon}(\omega_{k})
\overline{J_{\varepsilon}(\omega_{k+r})}
\exp\big(i(\phi(\omega_{k})- \phi(\omega_{k+r}))\big)  + O_{p}(T^{-1/2}),
\end{eqnarray*} 
where $J_{\varepsilon}(\omega) = 
(2\pi T)^{-1/2}\sum_{t=1}^{T}\varepsilon_{t}\exp(it\omega_{k})$.
\end{remark}

We use the martingale central limit theorem to show 
asymptotic normality of
$\sqrt{T}d_{T}(r)$, which will imply asymptotic normality of 
$\sqrt{T}\big(\tilde{c}_{T}(r) - \Ex(\tilde{c}_{T}(r)\big)$.
To do this we rewrite $\sqrt{T}d_{T}(r)$ as the 
sum of martingale differences
\begin{eqnarray*}
&&\sqrt{T}d_{T}(r) \\
&& = \frac{1}{T^{1/2}}\sum_{s_{1},s_{2}=1}^{T}
\big(\varepsilon_{s_{1}}\varepsilon_{s_{2}}
- \Ex(\varepsilon_{s_{1}}\varepsilon_{s_{2}}) \big)
\sum_{s_{1}\leq t\leq T}\sum_{s_{2}\leq \tau\leq T}
G_{T,\omega_{r}}(t-\tau)\exp(-i\tau\omega_{r})\psi_{t,T}(t-s_{1})\psi_{\tau,T}(\tau-s_{2}) \\
&& = \frac{1}{T^{1/2}}\sum_{s=1}^{T}M_{T}(s) \quad 
\textrm{ where }\quad 
M_{T}(s) = \big(\varepsilon_{s}^{2} -1\big)A_{T}(s,s) + 
\varepsilon_{s}\sum_{s_{1}=1}^{s-1}\varepsilon_{s_{1}}\big(A_{T}(s_{1},s) +A_{T}(s_{1},s) \big)
\end{eqnarray*}
and 
\begin{eqnarray*}
A_{T}(s_{1},s_{2}) = \sum_{s_{1}\leq t\leq T}\sum_{s_{2}\leq \tau\leq T}
G_{T,\omega_{r}}(t-\tau)\exp(-i\tau\omega_{r})\psi_{t,T}(t-s_{1})\psi_{\tau,T}(\tau-s_{2}). 
\end{eqnarray*}
We now show that the coefficients in the martingale differences
are absolutely summable. 
\begin{lemma}\label{lemma:moments}
Suppose Assumption \ref{assum:nonstat} holds. Then we have 
\begin{eqnarray*}
\sup_{T}\sum_{s_{1}=1}^{s-1}(|A_{T}(s,s_{1})| + |A_{T}(s_{1},s)|)  < \infty. 
\end{eqnarray*}
\end{lemma}
PROOF. To prove the result we note that 
under Assumption \ref{assum:nonstat} and using 
Lemma \ref{lemma:GDFT} we have
\begin{eqnarray*}
\sum_{s_{1}=1}^{s-1}|A_{T}(s_{1},s)| &\leq& \sum_{s_{1}=1}^{s-1}
\sum_{s_{1}\leq t\leq T}\sum_{s\leq \tau\leq T}
|G_{T,\omega_{r}}(t-\tau)|\cdot|\psi_{t,T}(t-s_{1})|\cdot|\psi_{\tau,T}(\tau-s)| \\
 &\leq& \bigg[\sum_{t}|G_{T,\omega_{r}}(t)|\bigg]\sup_{t,T}\bigg[\sum_{s}|\psi_{t,T}(s)|\big]^{2},
\end{eqnarray*}
which gives the required result. \hfill $\Box$

\vspace{3mm}

In the theorem below we show asymptotic normality of 
$d_{T}(r)$. To accommodate both the stationary and
nonstationary case we will let the asymptotic variance of 
$d_{T}(r)$ be $V_{r}$ and specify its value later. 

\begin{theorem}\label{thm:CLT1}
Suppose  Assumption \ref{assum:nonstat} holds. Furthermore suppose that 
$
\var(\sqrt{T}d_{T}(r), \sqrt{T}d_{T}(r)) 
\rightarrow V_{r}<\infty$
as $T\rightarrow \infty$. Then we have 
\begin{eqnarray*}
\sqrt{T}\left(
\begin{array}{c}
\Re \sqrt{T}d_{T}(r)  \\
\Im \sqrt{T}d_{T}(r) \\
\end{array}
\right) \Dcon \mathcal{N}\big(0, V_{r} \big).
\end{eqnarray*}
\end{theorem}
PROOF. We use the martingale central limit theorem 
to prove the result. We will show asymptotic normality of 
$\Re \sqrt{T}d_{T}(r)$. However, 
using the same method it straightforward 
to show asymptotic normality for all linear combinations of $\Re \sqrt{T}d_{T}(r)$ and 
$\Im \sqrt{T}d_{T}(r)$ and thus by the Cramer-Wold device to show asymptotic normality of the 
random vector 
$\big(\Re \sqrt{T}d_{T}(r),\Im \sqrt{T}d_{T}(r) \big)$. 
Let $M_{1,T} = \Re M_{T}(s)$. 
To apply the martingale central limit theorem we need to verify that
the variance of $T^{-1/2}\sum_{d=1}^{T}M_{1,T}(s)$ is finite (which is assumed),  
Lindeberg's condition is satisfied and 
$\frac{1}{T}\sum_{s=1}^{T}\Ex(M_{1,T}(s)^{2}|\mathcal{F}_{s-1}) 
\Pcon V_{r,1}$ (see \cite{b:hal-hey-80}, Theorem 3.2). 
To verify Lindeberg's condition, we require that for all $\delta > 0$, 
\begin{eqnarray*}
 L_{T} = \frac{1}{T}\sum_{s=1}^{T} \Ex(M_{1,T}(s)^{2}
I(T^{-1/2}|M_{1,T}(s)|>\delta)|\mathcal{F}_{s-1}) \Pcon 0,
\end{eqnarray*}
$T\rightarrow \infty$, where $I(\cdot)$ 
is the identity function and $\mathcal{F}_{s} = 
\sigma(M_{1,T}(s),M_{1,T}(s-1),\ldots,M_{1,T}(1))$. 
By using H\"older and Markov inequalities,  
we obtain a bound for the following $L_{T}$ 
\begin{eqnarray}
\label{eq:Ltbd}
L_{T}^{} \leq (T\delta)^{-1}\frac{1}{T}\sum_{s=1}^{T}\Ex(M_{1,T}(s)^{4}|\mathcal{F}_{s-1}). 
\end{eqnarray}
Now by using Lemma \ref{lemma:moments} we have 
$\sum_{s_{1}}\big(|A_{T}(s,s_{1})| +
A_{T}(s_{1},s)| \big)<\infty$, therefore 
\begin{eqnarray*}
\sup_{T}\Ex\big( \frac{1}{T}\sum_{s=1}^{T}
\Ex(M_{1,T}(s)^{4}|\mathcal{F}_{s-1}) \big) = 
\frac{1}{T}\sup_{T}\sum_{s=1}^{T}\Ex(M_{1,T}(s)^{4}) < \infty. 
\end{eqnarray*}
Since $\frac{1}{T}\sum_{s=1}^{T}\Ex(M_{1,T}(s)^{4}|\mathcal{F}_{s-1})$ 
is a positive random variable, the above result implies 
\\*
$\frac{1}{T}\sum_{s=1}^{T}\Ex(M_{1,T}(s)^{4}|\mathcal{F}_{s-1}) = O_{p}(1)$. 
Substituting this into (\ref{eq:Ltbd}) gives 
 $L_{T}\Pcon 0$ as $T\rightarrow \infty$. 

Finally we need to show that 
\begin{eqnarray}
\label{eq:nabla2}
\frac{1}{T}\sum_{s=1}^{T}\Ex(M_{1,T}(s)^{2}|\mathcal{F}_{s-1}) = 
\frac{1}{T}\sum_{s=1}^{T}\big[\Ex(M_{1,T}(s)^{2}|\mathcal{F}_{s-1}) - 
\Ex(M_{1,T}(s)^{2})\big] +\frac{1}{T}\sum_{s=1}^{T}\Ex(M_{1,T}(s)^{2})
\Pcon V_{r,1}. 
\end{eqnarray}
Under the stated assumptions,  we have 
$\frac{1}{T}\sum_{s=1}^{T}\Ex(M_{1,T}(s)^{2})\rightarrow V_{r,1}$
as $T\rightarrow \infty$. Therefore it remains to show  
\begin{eqnarray*}
P_{T} := \frac{1}{T}\sum_{s=1}^{T}
\big|\Ex(M_{1,T}(s)^{2}|\mathcal{F}_{s-1}) - \Ex(M_{1,T}(s)^{2})\big| \Pcon 0,
\end{eqnarray*}
which will give us (\ref{eq:nabla2}). We will show that $\Ex(P_{T}^{2})\rightarrow 0$. 
To do this we note that $\Ex(P_{T})=0$ and 
\begin{eqnarray}
\var(P_{T}) = \frac{1}{T^{2}}\sum_{d=1}^{T}\var(\Ex(M_{1,T}(s)^{2}|\mathcal{F}_{s-1})) + 
\frac{2}{T^{2}}\sum_{s_{1}>s_{2}}^{T}\cov(\Ex(M_{1,T}(s_{1})^{2}|\mathcal{F}_{s_{1}-1}), 
\Ex(M_{1,T}(s_{2})^{2}|\mathcal{F}_{s_{2}-1})). \label{eq:QQQd1} 
\end{eqnarray}
Now by using the Cauchy Schwartz inequality and conditional expectation
arguments for $\mathcal{F}_{s_{2}} \subset \mathcal{F}_{s_{1}}$ we have
\begin{eqnarray*}
&&\cov(\Ex(M_{1,T}(s_{1})^{2}|\mathcal{F}_{s_{1}-1}), 
\Ex(M_{1,T}(s_{2})^{2}|\mathcal{F}_{s_{2}-1})) \nonumber\\
&\leq &
\big[\Ex\big(\Ex(M_{1,T}(s_{2})^{2}|\mathcal{F}_{s_{2}-1}) - 
\Ex(M_{1,T}(s_{2})^{2})\big)^{2}\big]^{1/2}
\big[\Ex\big(\Ex(M_{1,T}(s_{1})^{2}|\mathcal{F}_{s_{2}-1}) - 
\Ex(M_{1,T}(s_{1})^{2})\big)^{2}\big]^{1/2}. \label{eq:QQQd}
\end{eqnarray*}
We now show that 
$\sup_{T}\sum_{s_{1}=s_{2}}^{T}\big[\Ex\big(\Ex(M_{1,T}(s_{1})^{2}|\mathcal{F}_{s_{2}-1}) - 
\Ex(M_{1,T}(s_{1})^{2})\big)^{2}\big]^{1/2}<\infty$. 
Let $\mathcal{G}_{s} = \sigma(\varepsilon_{s},\varepsilon_{s-1},\ldots)$, then it is clear that 
for all $s$, $\mathcal{F}_{s}\subset\mathcal{G}_{s}$. Therefore, we have 
$\Ex\big[\Ex(M_{1,T}(s_{1})^{2}|\mathcal{F}_{s_{2}-1})^{2}\big]\leq  
\Ex\big[\Ex(M_{1,T}(s_{1})^{2}|\mathcal{G}_{s_{2}-1})^{2}\big]$ which gives
\begin{eqnarray*}
\Ex\bigg[\Ex(M_{1,T}(s_{1})^{2}|\mathcal{F}_{s_{2}-1}) - \Ex(M_{1,T}(s_{1})^{2})\bigg]^{2} &=&
\Ex\big[\Ex(M_{1,T}(s_{1})^{2}|\mathcal{F}_{s_{2}-1})^{2}\big] - 
\big[\Ex(M_{1,T}(s_{1})^{2})\big]^{2} \\
&\leq &
\Ex\big[\Ex(M_{1,T}(s_{1})^{2}|\mathcal{G}_{s_{2}-1})^{2}\big] - 
\big[\Ex(M_{1,T}(s_{1})^{2})\big]^{2}. 
\end{eqnarray*}
Expanding $M_{1,T}(s_{1})$ in terms of $\{\varepsilon_{t}\}$ and using 
$\sup_{T,t}\sum_{j}|\psi_{t,T}(j)|<\infty$, it can be shown that 
$\Ex\big[\Ex(M_{1,T}(s_{1})^{2}|\mathcal{G}_{s_{2}-1})^{2}\big] -
\big[\Ex(M_{1,T}(s_{1})^{2})\big]^{2}\rightarrow 0$ as $s_{1}\rightarrow \infty$,
and 
\begin{eqnarray*}
\var(P_{T}) \leq \frac{1}{T}\sum_{s_{2}=1}^{T}
\sup_{s_{2},T}
\sum_{s_{1}=s_{2}}^{T}\big(\Ex\big[\Ex(M_{1.T}(s_{1})^{2}|\mathcal{G}_{s_{2}-1})^{2}\big]
- \big[\Ex(M_{s_{1}}^{2})\big]^{2}\big)^{1/2}<\infty. 
\end{eqnarray*}
Substituting the above into (\ref{eq:QQQd}) we have 
$\var(P_{T}) = O(T^{-1})$, hence we have shown (\ref{eq:nabla2}), 
and the conditions of the martingale central limit theorem 
are satisfied, giving the required result. \hfill $\Box$ 

\vspace{3mm}

{\bf Proof of Theorem \ref{thm:cCLT}} 
Using Lemma \ref{lemma:cov-diff} we have 
\begin{eqnarray*}
\sqrt{T}\bigg(\Re \widetilde{c}_{T}(r_{1}),\ldots,
\Im \widetilde{c}_{T}(r_{m})\bigg)
&=& 
\sqrt{T}\bigg(\Re \widehat{c}_{T}(r_{1}),\ldots,
\Im \widehat{c}_{T}(r_{m})\bigg) + \\
&& O_{p}\bigg( m
\bigg[
(b+\frac{1}{\sqrt{bT}}) + \big(\frac{1}{bT^{1/2}} + b^{2}T^{1/2}\big)
\sum_{n=1}^{m}\big(\frac{1}{|r_{n}|} + \frac{1}{T^{1/2}}\big) 
\bigg]
\bigg). 
\end{eqnarray*}
Lemma \ref{lemma:variance-stat}, implies that 
$T\var\bigg(\Re \widehat{c}_{T}(r_{1}),\ldots,
\Im \widehat{c}_{T}(r_{m})\bigg)\rightarrow 
\textrm{diag}(1+\frac{1}{2}\kappa_{4}\varphi(\frac{r_{1}}{T}),\ldots,
\textrm{diag}(1+\frac{1}{2}\kappa_{4}\varphi(\frac{r_{m}}{T}))$. Combining 
this with a similar proof to the proof of Theorem \ref{thm:CLT1},  
gives  
\begin{eqnarray}
\label{eq:normalitytildec}
\sqrt{T}\bigg(\frac{1}{1+\frac{1}{2}\kappa_{4}\varphi(\frac{r_{1}}{T})}
\Re \widetilde{c}_{T}(r_{1}),\ldots,
\frac{1}{1+\frac{1}{2}\kappa_{4}\varphi(\frac{r_{m}}{T})}\Im 
\widetilde{c}_{T}(r_{m})\bigg)
\Dcon 
\mathcal{N}(0,I_{2m}).
\end{eqnarray}
Finally, since 
$m(b+\frac{1}{\sqrt{bT}}) + \big(\frac{1}{bT^{1/2}} + b^{2}T^{1/2}\big)
\sum_{n=1}^{m}\big(\frac{1}{|r_{n}|} + \frac{1}{T^{1/2}}\big) \rightarrow 0$, 
using  (\ref{eq:normalitytildec}) we have (\ref{eq:cCLTb}).
\hfill $\Box$

\vspace{3mm}

{\bf PROOF of Theorem \ref{thm:nonstatclt}}. 
The proof is identical to the proof of Theorem \ref{thm:cCLT}. Hence we omit 
the details. \hfill $\Box$


\bibliographystyle{apacite}
\bibliography{test_stat_biblio}

\end{document}